\documentclass[american,aps,pra,reprint,floatfix,nofootinbib,superscriptaddress]{revtex4-2}
\usepackage[unicode=true,pdfusetitle, bookmarks=true,bookmarksnumbered=false,bookmarksopen=false, breaklinks=false,pdfborder={0 0 0},backref=false,colorlinks=false]{hyperref}
\hypersetup{colorlinks,linkcolor=blue,citecolor=red,urlcolor=cyan}
\usepackage{graphics,epstopdf,amsthm,amsmath,amssymb,mathptmx,colortbl,color,bm,framed,mathrsfs}
\usepackage{comment}
\usepackage{csquotes}
\usepackage[T1]{fontenc}
\usepackage[up]{subfigure}
\usepackage[braket, qm]{qcircuit}
\usepackage{physics}
\usepackage{commath}
\usepackage{mathtools}
\usepackage{float}
\usepackage{appendix}
\usepackage{amsthm}
\usepackage[linesnumbered, ruled,vlined]{algorithm2e}
\usepackage{dsfont}
\usepackage{nicefrac}
\usepackage[capitalise]{cleveref}
\makeatother

\newcommand\blfootnote[1]{%
  \begingroup
  \renewcommand\thefootnote{}\footnote{#1}%
  \addtocounter{footnote}{-1}%
  \endgroup
}

\usepackage[demo]{graphicx}
\usepackage{centernot}

\usepackage[normalem]{ulem}

\usepackage{csquotes}

\usepackage{xcolor}

\begin{document}

\title{Scalable Measurement Error Mitigation via Iterative Bayesian Unfolding}

\author{Siddarth Srinivasan$^*$}
\affiliation{Paul G. Allen School of Computer Science and Engineering, University of Washington, Seattle, WA 98195}

\author{Bibek Pokharel$^*$}

\affiliation{Department of Physics and Astronomy, University of Southern California, Los Angeles, CA, 90089, USA}
\affiliation{Center for Quantum Information Science $\&$ Technology, University of Southern California, Los Angeles, CA 90089, USA}

\author{Gregory Quiroz}
\affiliation{Johns Hopkins University Applied Physics Laboratory, Laurel, Maryland 20723}   
\affiliation{William H. Miller III Department of
Physics $\&$ Astronomy, Johns Hopkins University, Baltimore, Maryland 21218, USA}

\author{Byron Boots}
\affiliation{Department of Computer Science and Engineering, University of Washington, Seattle, WA 98195}

\date{\today}

\begin{abstract}

    Measurement error mitigation (MEM) techniques are postprocessing strategies to counteract systematic readout errors on quantum computers (QC). Currently used MEM strategies face a tradeoff: methods that scale well with the number of qubits return negative probabilities, while those that guarantee a valid probability distribution are not scalable. Here, we present a scheme that addresses both of these issues. In particular, we present a scalable implementation of iterative Bayesian unfolding, a standard mitigation technique used in high-energy physics experiments. We demonstrate our method by mitigating QC data from experimental preparation of Greenberger-Horne-Zeilinger (GHZ) states up to 127 qubits and implementation of the Bernstein-Vazirani algorithm on up to 26 qubits.
\end{abstract}
\maketitle

\section{Introduction}

\blfootnote{$^*$ denotes equal contribution.}

Measurement errors are a significant source of error \cite{IBMQuantum2022,RigettiComputing2022,wrightBenchmarking11qubitQuantum2019,HoneywellQuantum2022} for today's programmable quantum computers (QCs). Currently, QCs typically implement quantum algorithms with a measurement at the end of the quantum circuit. However, advanced quantum operations, including quantum error correction~\cite{nielsenQuantumComputationQuantum2010a,Lidar-Brun:book}, involve measurements not only at the end of quantum computation but throughout the computational scheme. Thus, measurement accuracy is vital 
to obtaining accurate computational results. Measurement error mitigation (MEM) techniques are a collection of classical post-processing techniques that aim to address the measurement errors inherent to the QC hardware. Most MEM methods are unable to achieve both scalability and a guarantee of valid probability distributions after mitigation. Tackling this challenge, we present a scalable MEM strategy whose time and memory costs do not scale exponentially with the number of qubits while also guaranteeing valid probability distributions.

Most MEM methods assume that the true classical probability distribution $\vec{\theta}$ obtained from measuring a quantum state in the computational basis is modified by a response matrix $R$ to yield  $\vec{p}$, a noisy probability distribution distorted by measurement occurring in a faulty basis, i.e., 
\begin{equation}
    \vec{p} = R \vec{\theta}.
    \label{eq:pRt}
\end{equation}
\cref{eq:pRt} assumes that measurement errors are systematic, linear, and time-independent. To obtain the response matrix, most MEM techniques further assume that the state preparation errors can be ignored. These simplifying assumptions have been verified for IBM Quantum Experience (IBMQE) superconducting devices \cite{maciejewskiMitigationReadoutNoise2020,nationScalableMitigationMeasurement2021}, provided the readout calibration data is up to date.  

\begin{figure}
 \begin{displaymath}
\large
\Qcircuit @C=1em @R=0.9em @!R {
 &  \mbox{Quantum Detector Tomography} &   \\
\lstick{| 0 \rangle}  & \multigate{4}{\text{State Preparation}} & \meter  \\
\lstick{| 0 \rangle}  & \ghost{\text{State Preparation}}        & \meter  \\
                      & \ghost{\text{State Preparation}} 	& \vdots  \\
\lstick{| 0 \rangle}  & \ghost{\text{State Preparation}} 	& \meter  \\
\lstick{| 0 \rangle}  & \ghost{\text{State Preparation}} 	& \meter  
}
    \end{displaymath}

    \begin{displaymath}
\large
\Qcircuit @C=1em @R=0.9em @!R {
 & & & \text{Error-mitigated experiment} &  &   \\
\lstick{| 0 \rangle}  & \multigate{4}{\text{U}} & \meter  & \cw	   & \cw & \cw &  & & \cw		\\
\lstick{| 0 \rangle}  & \ghost{\text{U}}        & \meter  & \cw    & \cw & \cw & &  & \cw      		\\
                      & \ghost{\text{U}} 	&   	  & \vdots & 	 & & \push{\text{MEM}} & & \cw		\\
\lstick{| 0 \rangle}  & \ghost{\text{U}} 	& \meter  & \cw    & \cw & \cw & &  & \cw       		\\
\lstick{| 0 \rangle}  & \ghost{\text{U}} 	& \meter  & \cw    & \cw & \cw & &  & \cw \gategroup{2}{6}{6}{8}{1.2em}{--} \\}
    \end{displaymath}
    \caption{\emph{Two steps of MEM.} The quantum detector tomography (QDT) circuit is shown on top. QDT is performed to estimate the response matrix $R$. The bottom figure shows the schematic circuit for the MEM application. After applying a unitary $U$, the classical measurement data is fed into a MEM scheme to produce an error-mitigated probability distribution.   }
    \label{fig:mem-setup}
\end{figure}
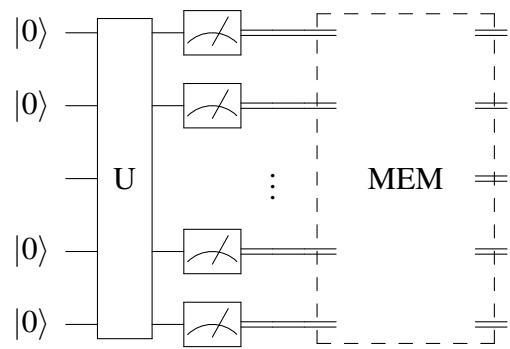

Assuming that systematic measurement errors manifest as in \cref{eq:pRt}, MEM schemes require two steps. The first step is quantum detector tomography (QDT) \cite{maciejewskiMitigationReadoutNoise2020} to acquire the response matrix $R$ (see \cref{fig:mem-setup}).  The entries of the response matrix  $R_{ij}$  are conditional probabilities, i.e., the probability of measuring a bitstring $b_i$ given that computational basis element $| b_j \rangle$ was prepared. In the second step, the classical probability distribution $\vec{p}$ acquired from measurements after running quantum circuit $U$ is used to compute a new probability distribution $\vec{\theta}$ which approximates the true distribution (for any future experiment). The most popular strategy, response matrix inversion, defines 
\begin{equation}
    \vec{\theta} = R^{-1} \vec{p}.
\end{equation}

Acquiring $R$ and implementing MEM face two challenges: \emph{scaling} and \emph{negative probabilities}. First, $R$ is a $2^n \times 2^n$ matrix for an $n$-qubit system and in general requires $2^n$ calibration experiments. In other words, the dimensions of $R$ are exponential in $n$, and QDT requires exponentially many experiments. Secondly, while the response matrix $R$ is stochastic, its inverse is not. As most MEM methods rely on applying $R^{-1}$, this leads to a `negative probability' issue. The distributions obtained after measurement error mitigation are \emph{quasiprobabilities} - the elements of this quasiprobability distribution can be negative or greater than 1 as long as they sum to 1.  

Various strategies have been proposed to address the scalability of the response matrix inversion method. Assuming $k$-locality for measurement crosstalk \cite{bravyiMitigatingMeasurementErrors2021, maciejewskiMitigationReadoutNoise2020} allows $R$ to be represented as $\nicefrac{n}{k}$ different $2^k \times 2^k$ matrices. This assumption has been verified in Refs~\cite{nationScalableMitigationMeasurement2021, maciejewskiMitigationReadoutNoise2020} and presents the first step in addressing the scalability issue. The unmitigated probability distributions are already sparse as the experiments required to construct these distributions are only performed for $S$ shots, which are generally set between 500 to 100000. In practice, $S$ does not scale exponentially with qubit size due to time and computational constraints. Ref~\cite{nationScalableMitigationMeasurement2021} relied on this sparsity to make MEM tractable. In particular, only the matrix elements of $R$ that had bitstrings close to the observed ones (measured using Hamming distance) could be considered, while others were ignored. This \emph{subspace reduction} of $R$ suggested that a user could perform QDT after and with the knowledge of the unmitigated experimental results. In other words,  the order of QDT and the experiment in \cref{fig:mem-setup} was flipped. Similarly, Ref.~\cite{mooneyGenerationVerification27qubit2021} suggested placing a precision threshold on $R$ and MEM, such that any probabilities below $\nicefrac{1}{(10\times S)}$ are ignored. These approaches to scaling MEM have been quite effective. In particular, Ref.~\cite{mooneyGenerationVerification27qubit2021, nationScalableMitigationMeasurement2021, yangEfficientQuantumReadout2022} have demonstrated Greenberger-Horne-Zeilinger (GHZ)  state preparation on 27, 42 and 65-qubits respectively. MEM plays a crucial part in these demonstrations by significantly improving success metrics.

The methods listed above, while scalable, aim to invert $R$ and apply $R^{-1}$, and the non-stochasticity of $R^{-1}$ necessarily leads to quasiprobabilities. Numerous methods~\cite{IBMQuantum2022,nachmanUnfoldingQuantumComputer2020,smolinEfficientMethodComputing2012} have been proposed to address the negativity issue, but no clear consensus has emerged.  One strategy is to use constrained least-squares optimization to find a probability distribution with no negative probabilities close to the mitigated quasiprobability distribution (in terms of some norm-based distance between probability distributions). However, it is unclear how to scale this approach. Alternatively, the negative terms can be canceled systematically by projecting to the nearest valid probability distribution, as detailed in Ref.~\cite{smolinEfficientMethodComputing2012}. However, this zeroes out all negative probabilities and can no longer be considered an unbiased estimator of the true distribution without measurement errors. Here Ref.~\cite{nachmanUnfoldingQuantumComputer2020} is particularly notable. It highlighted that error-prone detectors are standard across experimental fields, and unfolding techniques used in high-energy physics experiments can also be used to mitigate readout errors on quantum devices. In particular, Ref.~\cite{nachmanUnfoldingQuantumComputer2020} showed that iterative Bayesian unfolding (IBU)~\cite{dagostiniMultidimensionalUnfoldingMethod1995} could mitigate readout errors while preserving non-negativity.

While IBU does not return any quasi-probabilities in its basic form, it requires iterative matrix multiplications with $R$ and is challenging to scale to many qubits without further modification. In this paper, we demonstrate a scalable implementation of IBU. The original IBU formulation starts with the response matrix $R$, the noisy empirical distribution $\vec{p}$, and an initial guess $\vec{\theta}^0$. Then it iteratively applies Bayes' rule to find a mitigated probability distribution $\vec{\theta}^k$ like so:
\begin{equation}\label{eq:ibu}
    {\theta}_{j}^{k+1} = \sum_{i=1}^{2^n} p_i  \cdot \frac{R_{ij} {\theta}_j^{k}}{\sum_{m} R_{im}  {\theta}^{k}_{m}}
\end{equation}

A well-known technique in the high-energy physics community, IBU is also known as the Richardson-Lucy deconvolution~\cite{richardsonBayesianBasedIterativeMethod1972,lucyIterativeTechniqueRectification1974}. 
While IBU uses Bayes' rule and starts with an initial guess, sometimes referred to as a `prior,' it does not report a `posterior' in any sense and is not a Bayesian method. It is more appropriate to call it iterative expectation maximization unfolding~\cite{volobouevExpectationMaximizationUnfoldingSmoothing2015} as it converges to the maximum likelihood estimate~\cite{sheppMaximumLikelihoodReconstruction1982} for a large number of iterations. Since this method may not be familiar to the quantum computing community,  we provide a derivation of IBU as expectation maximization as relevant to the quantum computing setting.

Our main challenge was to show that under reasonable assumptions about measurement crosstalk and precision, IBU can be implemented scalably. To demonstrate the scalability of our method, we mitigate errors on QC data acquired by implementing the Bernstein-Vazirani algorithm~\cite{bernsteinQuantumComplexityTheory1997} on the 27-qubit IBMQ Montreal~\cite{pokharelDemonstrationAlgorithmicQuantum2022} and from preparing GHZ states for the 127-qubit IBMQ Washington. Our MEM implementation performs as well as or better than the current state-of-the-art M3 method \cite{nationScalableMitigationMeasurement2021} \emph{without} producing negative probabilities, albeit with slightly higher computational time. 




The rest of the paper is structured as follows: in \cref{sec:ibu}, we describe our implementation of scalable iterative Bayesian unfolding. We discuss the connection between IBU and expectation maximization in \cref{sec:em}. \cref{sec:demonstrations} focuses on demonstrating our MEM method on two quantum datasets -- the Bernstein-Vazirani algorithm on up to 26-qubits (\cref{sec:bv}) and GHZ state preparation up to 81 qubits (\cref{sec:ghz}). We give some concluding remarks in \cref{sec:conclusion}.



\section{Scalable IBU Implementation}
\label{sec:ibu}

Here, we discuss some assumptions and computational approaches we use to scalably implement IBU (as in Eq. \ref{eq:ibu}) for measurement error mitigation (see Algorithm \ref{alg:ibu}).

\paragraph{Vectorized Implementation} First, we show how to vectorize the IBU update rule given in Equation \ref{eq:ibu} to enable fast parallel computation on GPUs. Let $Z$ be a random variable representing the true distribution over bitstrings without any measurement error and $Z'$ be a random variable representing the distribution over bitstrings with measurement error. With $c(z'_i)$ the counts of bitstring $z'_i$ and $S$ shots, let $\vec{p} \in \mathds{R}^{2^n}$ be the noisy distribution with $p_i = \frac{c(z'_i)}S$, let ${\vec{\theta}^k} \in \mathds{R}^{2^n}$ be the $k$th iteration guess of the error-mitigated distribution with ${{\theta}^k}_j = P_{\vec{\theta}^k}(Z_j)$, and let $R \in \mathds{R}^{2^n \times 2^n}$ be the response matrix where $R_{ij} = P(Z'_i | Z_j)$. Then, the IBU update rule is:


\begin{equation}\label{eq:ibu4}
\begin{split}
    \vec{\theta}^{k+1} &= \sum_{i=1}^{2^n} p_i  \cdot \frac{R_i \odot {\vec{\theta}^k}}{R_i {\vec{\theta}^k}} \\
    &= \left(\vec{p}^T \cdot  \text{diag}\left(\frac1{R{\vec{\theta}^k}}\right) \cdot R \cdot \text{diag}({\vec{\theta}^k})\right)^T \\
    &= {\vec{\theta}^k} \odot \left(R^T\left(\vec{p} \oslash R{\vec{\theta}^k}\right)\right)
\end{split}
\end{equation}

Here, $\odot$ is element-wise multiplication, and $\oslash$ is elementwise division. This expression fully vectorizes the IBU update rule by writing it solely in terms of elementwise operations and matrix products. It also optimizes the order of operations to avoid constructing large intermediate matrices; only the additional memory for storing $\vec{\theta}^{k+1}$ is needed.

\paragraph{Tensor Product Structure of Noise Model} Despite the efficient update scheme above, the memory requirements are still quite unfavorable. As previously discussed, the first big memory bottleneck is that the response matrix $R$ grows exponentially in the number of qubits: $2^n \times 2^n$. To get around this issue, we assume that there is no measurement crosstalk and hence $R$ has a convenient tensor product decomposition \citep{maciejewskiMitigationReadoutNoise2020, bravyiMitigatingMeasurementErrors2021, nationScalableMitigationMeasurement2021}, i.e., $R = R^{(1)} \otimes R^{(2)} \otimes \ldots \otimes R^{(n)}$. While we restrict ourselves to this scenario, our implementation can be extended to cases where measurement crosstalk is restricted to $k$ qubits. 

With this tensor product structure on $R$, we now have $4n$ parameters to represent $R$, as opposed to the original $4^n$ parameters. This structure also allows matrix product $R\vec{x}$ to be computed very quickly using only matrix products and reshapes \citep{fackler2019algorithm} without any additional memory requirement. This is the simplest structure we can impose on $R$ to obtain tractability, although this can be relaxed into other tensor network decompositions of $R$ as well (such as matrix product states (MPS)).

\paragraph{Subspace Reduction for Tractability}
While we managed the exponential scaling of $R$ with a tensor product decomposition, the parameters of the true latent distribution $\vec{\theta} \in \mathds{R}^{2^n}$ also scale exponentially in the number of qubits. To manage this, we can use the subspace reduction used in Ref.~\citep{nationScalableMitigationMeasurement2021}. In particular, instead of maintaining and updating a vector of probabilities over $2^n$ bitstrings, we only maintain and update an $M$-dimensional vector over select bitstrings. Specifically, we pre-select $M$-bitstrings (e.g., to be all those bitstrings within Hamming distance $d$ from any bitstring in our dataset) and initialize $\vec{\theta}^0$ with non-zero values in the entries corresponding to those $M$ bitstrings; all other entries are zero. Under such an initialization $\vec{\theta}^0$, the IBU update rule will guarantee that all entries that started as zero remain zero. Thus, we can ignore those entries and track only the $M \times 1$ sub-vector corresponding to non-zero entries. We refer to this subspace-reduced implementation as \emph{IBU (Reduced)}, and IBU without subspace reduction as \emph{IBU (Full)}. Under this subspace reduction trick, we can no longer easily leverage the Kronecker product structure of $R$ to compute $R\vec{x}$ for any $\vec{x}$. Instead, we must construct a reduced matrix $\tilde{R}$ whose rows and columns correspond to the $M$ selected bitstrings. Unlike \citep{nationScalableMitigationMeasurement2021}, we are not solving a least squares problem, so matrix-free methods like generalized minimal residual method (GMRES) will not be helpful; we need to construct the reduced matrix $\tilde{R}$. Even if the matrix does not grow exponentially with the number of qubits, it may still be prohibitively large to store in memory. Our computational solution to this problem is to build the solution to $R\vec{x}$ by weighting each column $R_j$ by $x_j$ and only keeping the running sum. We also use the JAX library's built-in accelerated linear algebra routines and just-in-time compilation, vectorize as many operations as possible, and use a GPU for parallelization. Although we incur the cost of repeatedly computing $\tilde{R}$, there is a tradeoff between this cost and memory costs for explicitly storing ${R}$. 

\paragraph{Worst-case Computational Complexity} Suppose we run an experiment with $S$ shots on an $n$ qubit system and track strings up to Hamming distance $d$ from the measured bitstrings. In the worst case, there will be $S$ different measurements, so subspace-reduced IBU tracks $S\cdot\left(\sum_{k=0}^d {n \choose k}\right) \sim O(Sn^d)$ bitstrings. The three main computational subroutines are 1) constructing the subspace-reduced response matrix, 2) matrix-vector multiplication, and 3) elementwise multiplication and division. First, the subspace-reduced response matrix will have $O(S^2n^{2d})$ cells to be constructed. Each cell of the matrix requires a single call to each of the $n$ single-qubit response matrices, so constructing the matrix takes $O(S^2n^{2d+1})$ operations. Second, multiplying this matrix with a vector with $O(Sn^d)$ entries takes $O(S^2n^{2d})$ operations. Third, the elementwise multiplication and division operations take $O(Sn^d)$ operations. Overall, a single IBU iteration scales as $O(S^2n^{2d+1})$, better than exponential scaling of the naive approach.

\begin{algorithm}[h
    ]\label{alg:ibu}
    \caption{Scalable IBU}
\SetAlgorithmName{Algorithm}{}
\DontPrintSemicolon
\KwIn{$\vec{p}$ (Vector of normalized counts for each bitstring), $\{R^{(i)}\}_{i=1}^n$ ($n$ single-qubit response matrices), $\vec{\theta}^0$ (initial guess of MEM distribution), tol (the convergence tolerance)} 
\KwOut{$\vec{\theta}^k$ (MEM distribution) } 

\While{$\|\vec{\theta}^{k+1} - \vec{\theta}^k \| < \text{tol}$}{

Compute $\vec{x}_1 = ({R^{(1)}} \otimes R^{(2)} \otimes \ldots \otimes R^{(n)}){\vec{\theta}^k}$ using \citep{fackler2019algorithm} if no subspace reduction, else by keeping a running sum of the columns $R_j$ weighted by $\vec{\theta}^k_j$ \\

Compute $\vec{x}_2 = \vec{p} \oslash \vec{x}_1$ \\

Compute $\vec{x}_3 = R^T \vec{x}_2$ using \citep{fackler2019algorithm} if no subspace reduction, else by keeping a running sum of the columns $R^T_j$ weighted by $(\vec{x}_2)_j$ \\

Compute $\vec{\theta}^{k+1} = \vec{\theta}^k \odot \vec{x}_3$.

}
Return $\vec{\theta}^k$
\end{algorithm}

\section{Iterative Bayesian Unfolding as Expectation-Maximization}
\label{sec:em}
Here we summarize the argument that iterative Bayesian unfolding is an instantiation of the well-known \emph{Expectation-Maximization} (EM) algorithm from machine learning. See Appendix \ref{sec:fullderiv} for complete derivation as well as a `true' Bayesian extension that computes that maximum a posteriori (MAP) estimate of $\vec{\theta}$ given Dirichlet priors.  \citet{sheppMaximumLikelihoodReconstruction1982} have noted connections between IBU and EM for Poisson variables. In our setting, we have at most $2^n$ discrete observations, each with an associated probability. So our measurements come from a multinomial distribution parameterized by $\vec{\theta} \in \mathds{R}^{2^n}$, and we show the IBU-EM connection for multinomial distributions typical in quantum computing. 

EM is a standard maximum-likelihood approach to estimating the parameters of latent variable models, where optimizing the likelihood analytically and directly is not possible. The EM algorithm iterates between an \emph{E-step}, which estimates a conditional expected likelihood under a given choice of model parameters, and an \emph{M-step} which updates the model parameters to maximize the aforementioned conditional likelihood. EM maximizes a lower bound of the likelihood at every iteration, and thus every parameter update increases the likelihood. The method is guaranteed to converge to a local maximum. 

\paragraph{Setup} Assume all measurements are made in a fixed basis. Let $Z$ be the discrete random variable denoting $n$-length bitstrings with $P(Z)$ be the distribution over bitstrings prepared by the quantum computer. Due to measurement error, the distribution over bitstrings that are actually measured is different, and we denote this with random variable $Z'$ with distribution $P(Z')$. Suppose we collect a dataset of $S$ shots ${\bf Z}' = \{z'_1, \ldots, z'_S\}$ sampled IID from $P(Z')$, and that we have access to the error model $P(Z'|Z)$. Our goal is to estimate $P_\theta(Z)$, where $\vec{\theta}$ is $2^n$-dimensional vector summing to 1 that parameterizes the true distribution over bitstrings. 

\paragraph{Expectation-Maximization} Using Jensen's inequality, we can manipulate the log-likelihood of the data ${\bf Z'}$ under parameter $\vec{\theta}$ as:
\begin{equation}\small \label{eq:bound}
\begin{split}
    \ell(\vec{\theta}; {\bf z}') &= \sum_{i=1}^S \log \left(\sum_{j=1}^{2^n} P_\theta(z_j) P(z'_i | z_j)  \right) \\
    &= \sum_{i=1}^S \log \left(\sum_{j=1}^{2^n} P_{\theta^k}(z_j|z'_i) \frac{P(z'_i | z_j) P_\theta(z_j)}{P_{\theta^k}(z_j|z'_i)} \right) \\
    &\geq \sum_{i=1}^S \sum_{j=1}^{2^n} P_{\theta^k}(z_j|z'_i) \log \left( \frac{P(z'_i, z_j; \vec{\theta})}{P_{\theta^k}(z_j|z'_i)} \right) \equiv \hat{\ell}(\vec{\theta} | \vec{\theta}^k; {\bf z}')
    \end{split}
\end{equation}

We can then find $\vec{\theta}$ that maximizes this lower bound $\hat{\ell}(\vec{\theta} | \vec{\theta}^k; {\bf Z'})$ locally, which is also equivalent to maximizing $\ell(\vec{\theta} | \vec{\theta}^k; {\bf Z'}) := \sum_{i=1}^S \sum_{j=1}^{2^n} P_{\theta^k}(z_j|z'_i) \log P(z'_i, z_j; \vec{\theta}) $ since the denominator inside the log in $\hat{\ell}(\vec{\theta}, \vec{\theta}^k; {\bf Z'})$ does not depend on $\vec{\theta}$, i.e.,

\begin{equation}
        \vec{\theta}^{k+1} = \underset{\vec{\theta}}{\text{arg max }} \hat{\ell}(\vec{\theta} | \vec{\theta}^k; {\bf Z'}) = \underset{\vec{\theta}}{\text{arg max }} \ell(\vec{\theta} | \vec{\theta}^k; {\bf Z'})  
\end{equation}

To locally maximize $\ell(\vec{\theta} | \vec{\theta}^k; {\bf Z'})$, we can take the derivative (using the Lagrange multiplier $\lambda$ to enforce the normalization constraint that $\sum_j {\theta}_j = 1$).

\begin{equation} \small
        \phantom{=} \frac{\partial }{ \partial {\theta}_j} \left( \ell(\vec{\theta} | \vec{\theta}^k; {\bf Z'})  + \lambda \left(1 - \sum_{j'=1}^{2^n} {\theta}_{j'} \right) \right) = \left(\sum_{i=1}^S P_{\theta^k}(Z_j|Z'_i) \right)\frac1{{\theta}_j} - \lambda
\end{equation}

Setting this derivative to zero, we find that $\lambda = S$. Substituting $\lambda= S$ back into the derivative set to zero, we get the update rule for each entry ${\theta}_j$ of $\vec{\theta}$ that locally maximizes the log-likelihood given guess $\vec{\theta}^k$:

\begin{equation}
    {\theta}^{k+1}_j = \frac1S \sum_{i=1}^S P_{\theta^k}(z_j|z'_i)
\end{equation}

Using Bayes' rule, we have $ P_{\theta^k}(z_j|z'_i) = \frac{P(z'_i | z_j) P_{\theta^k}(z_j)}{\sum_m P(z'_i | z_m) P_{\theta^k}(z_m)}$. Instead of summing over every bitstring in the dataset, we can re-write the sum as over all possible bitstrings using bitstring counts $c(z'_i)$. Doing so, we end up with the IBU update rule:

\begin{equation}\label{eq:ibu3}
\begin{split}
    {\theta}^{k+1}_j &= \sum_{i=1}^{2^n} \frac{c(z'_i)}S  \cdot \frac{P(z'_i | z_j) P_{\theta^k}(z_j)}{\sum_m P(z'_i | z_m) P_{\theta^k}(z_m)} \\
    &= \sum_{i=1}^{2^n} p_i  \cdot \frac{R_{ij} \theta_j^{k}}{\sum_m R_{im}  \theta^{k}_m}
    \end{split}
\end{equation}


\paragraph{Convergence} Finally, we note that EM converges to a local stationary point of the log-likelihood. This is because the optimization objective $\hat{\ell}(\vec{\theta} | \vec{\theta}^k; {\bf Z'})$ lower bounds the log-likelihood of the data ${\ell}(\vec{\theta}; {\bf Z'})$, and is also equal to the log-likelihood at $\vec{\theta} = \vec{\theta}^k$. 
Consequently, 
any choice of $\vec{\theta}^{k+1}$ which improves $\hat{\ell}(\vec{\theta}^{k+1} | \vec{\theta}^k; {\bf Z'})$ from the current estimate $\hat{\ell}(\vec{\theta}^k | \vec{\theta}^k; {\bf Z'})$ must also improve the log-likelihood $\ell(\vec{\theta}^k; {\bf Z'})$. This log-likelihood increases at every iteration of IBU by at least the amount $\hat{\ell}(\vec{\theta}^{k+1} | \vec{\theta}^k; {\bf Z'}) - \hat{\ell}(\vec{\theta}^k | \vec{\theta}^k; {\bf Z'})$ and the iterations stop at a stationary point of $\hat{\ell}(\vec{\theta} | \vec{\theta}^k; {\bf Z'})$. Since the optimization objective $\hat{\ell}(\vec{\theta} | \vec{\theta}^k; {\bf Z'})$ and log-likelihood $\ell(\vec{\theta}; {\bf Z'})$ are equal at $\vec{\theta} = \vec{\theta}^k$ and have the same slope with respect to $\vec{\theta}$ \citep{bishop2006pattern}, the termination coincides with a stationary point of the log-likelihood.

\section{Demonstrations}
\label{sec:demonstrations}
We demonstrate our method on two datasets -- measurements after the Bernstein-Vazirani (BV) algorithm (26 qubits) and measurements of the GHZ state (up to 127 qubits). We compare our method with M3 and focus on three metrics: accuracy (measured as success probabilities for BV and $\ell_1$-error for GHZ), runtime on our machine, and total negative probabilities in the mitigated distribution. Our algorithm is implemented using JAX 0.3.10 \citep{jax2018github} and run on an NVIDIA GeForce RTX 3090 GPU with 24576 MB of memory. \footnote{Our code is accessible at: \texttt{https://github.com/sidsrinivasan/PyIBU}.}

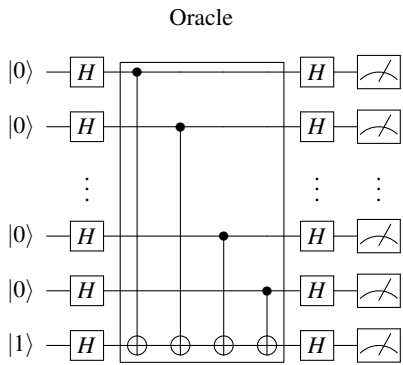
\begin{figure}[!htpb]
    \centering
    \vspace{-5mm}
        \begin{displaymath}
    \Qcircuit @C=1em @R=0.9em @!R {
     & & &  \mbox{\ \ \ \ \ \ \ Oracle} &  &  \\
    \lstick{| 0 \rangle} & \gate{H} & \ctrl{5} & \qw & \qw & \qw & \gate{H} & \meter  \\
    \lstick{| 0 \rangle} & \gate{H} & \qw & \ctrl{4} & \qw & \qw & \gate{H} & \meter  \\
    & \vdots &  & & & & \vdots & \vdots  \\
    \lstick{| 0 \rangle}  & \gate{H} & \qw & \qw & \ctrl{2} & \qw & \gate{H} & \meter  \\
    \lstick{| 0 \rangle} & \gate{H} & \qw & \qw & \qw & \ctrl{1} & \gate{H} & \meter  \\
    \lstick{| 1 \rangle}  & \gate{H}  & \targ & \targ & \targ & \targ & \gate{H} & \meter \gategroup{2}{3}{7}{6}{0.6em}{-}
    }
        \end{displaymath}
    \caption{\emph{Bernstein-Vazirani circuit.} Circuit diagram for the Bernstein-Vazirani algorithm with the hidden bitstring $b=1\dots1$~\cite{bernsteinQuantumComplexityTheory1997, nielsenQuantumComputationQuantum2010a}.
    \label{fig:bv_circuit}}
    
    \end{figure}
    
\begin{figure}[!htbp]
\begin{displaymath}
        \Qcircuit @C=1.0em @R=0.2em @!R { \\
     \lstick{{q}_{0}} & \qw & \qw & \targ & \qw & \qw & \qw & \meter\\
    \lstick{{q}_{1}} & \gate{\mathrm{H}} & \ctrl{2} & \ctrl{-1} & \ctrl{1} & \qw & \qw & \meter\\
    \lstick{{q}_{2}} & \qw & \qw & \qw & \targ & \qw & \qw & \meter\\
    \lstick{{q}_{3}} & \qw & \targ & \ctrl{2} & \qw & \qw & \qw & \meter &   & &  & & \includegraphics[trim = 70 70 70 70,clip,width=0.4\columnwidth]{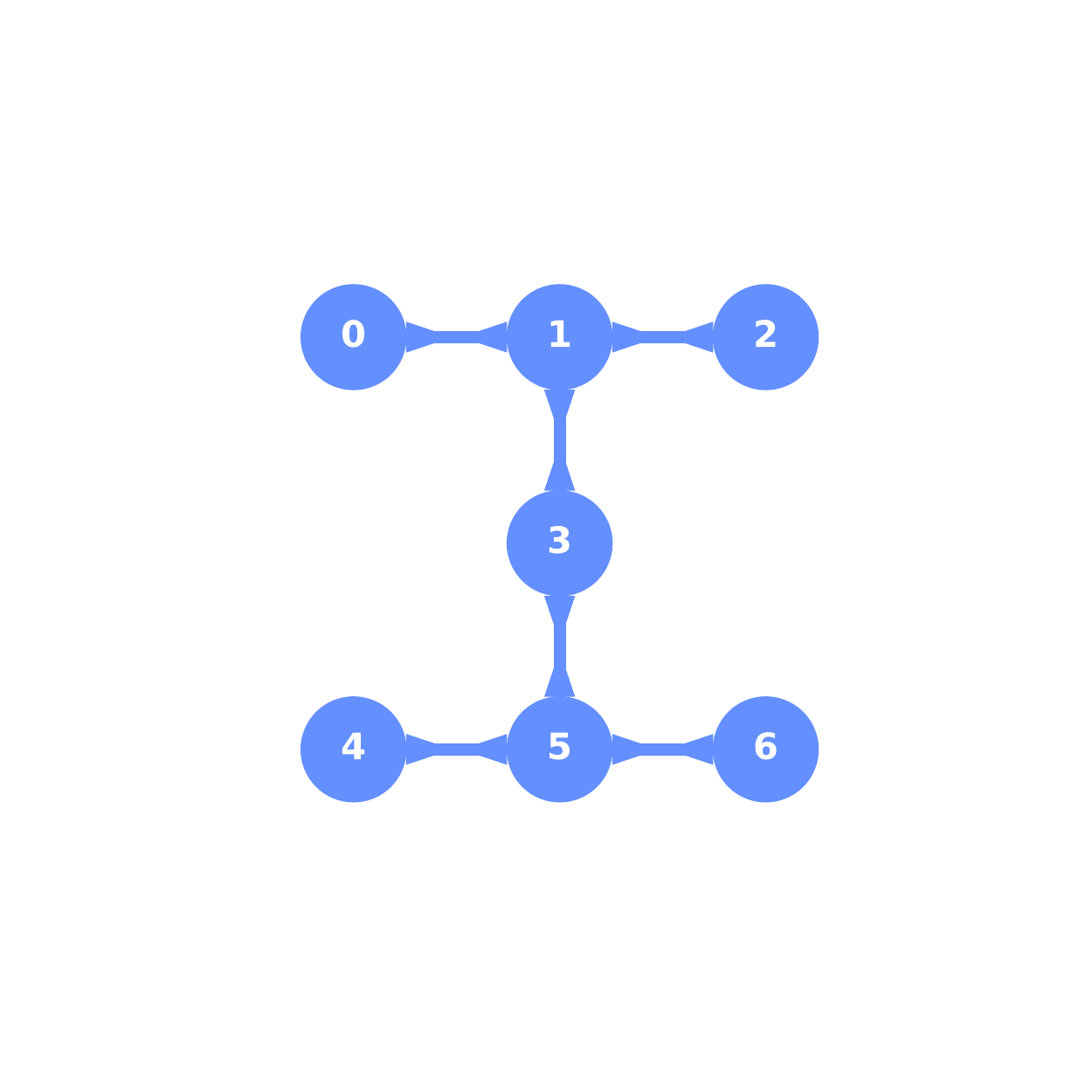}\\
    \lstick{{q}_{4}} & \qw & \qw & \qw & \targ & \qw & \qw & \meter\\
    \lstick{{q}_{5}} & \qw & \qw & \targ & \ctrl{-1} & \ctrl{1} & \qw & \meter\\
    \lstick{{q}_{6}} & \qw & \qw & \qw & \qw & \targ & \qw & \meter 
             }
     \end{displaymath}

        \caption{\emph{Example of GHZ generation circuit.} Given a 7-qubit architecture (right), the GHZ generation circuit is created by first identifying an vertex with maximum degree ($q_1$) and then using breadth-first expansion and performing CNOTs accordingly to create a fully-entangled state. For our experiments, we generated GHZ on an 127-qubit device, but here we demonstrate the scheme on a 7-qubit device for simplicity.}
        \label{fig:ghz-prep}
\end{figure}

\subsection{Bernstein-Vazirani algorithm}
\label{sec:bv}

\begin{figure*}[htpb!]
    \centering
    \begin{minipage}{0.6\linewidth}
            \begin{subfigure}{}
                \includegraphics[trim = 50 0 100 100, clip,width=\textwidth]{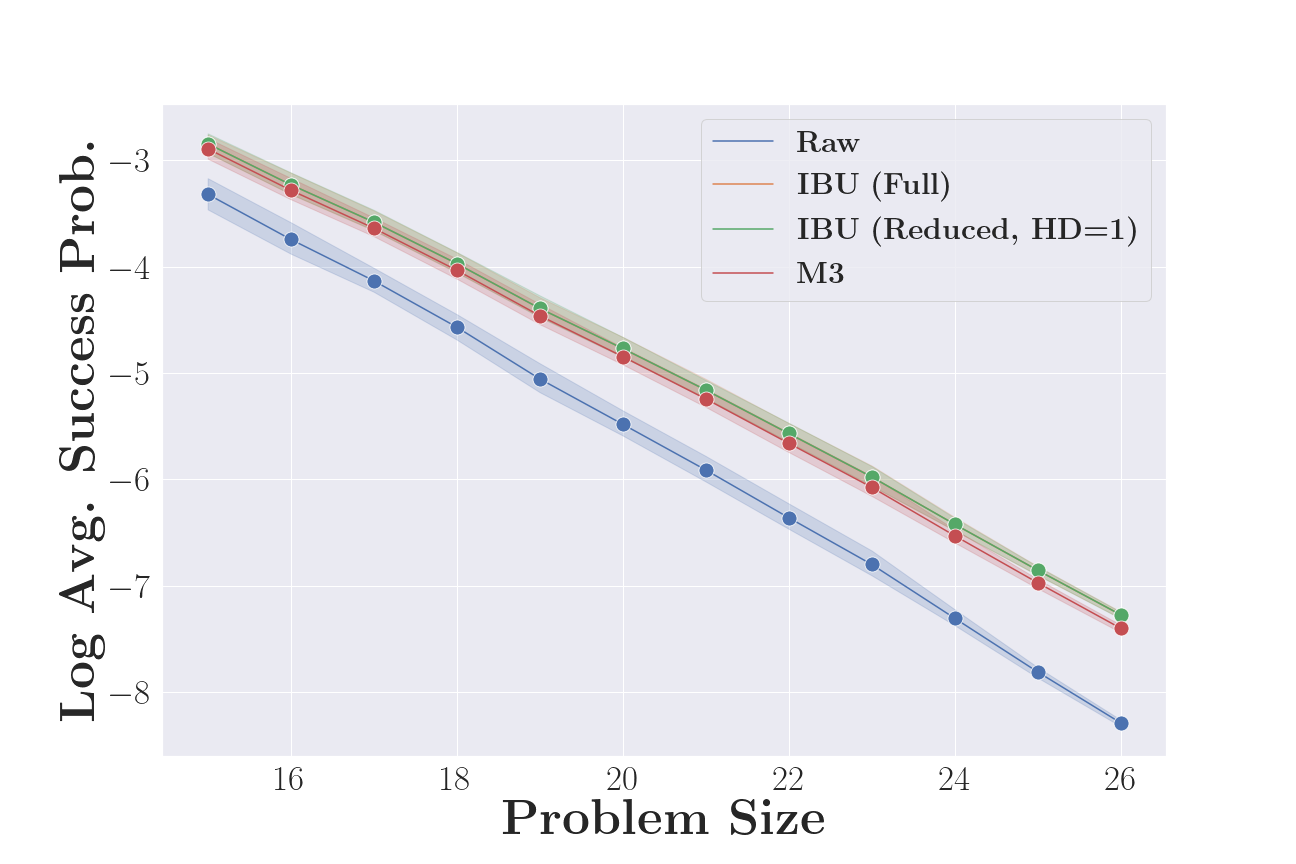}
            \end{subfigure}
        \end{minipage}
    \begin{minipage}{0.38\linewidth}
        \begin{subfigure}{}
            \includegraphics[trim = 0 70 100 0, clip, width=\textwidth]{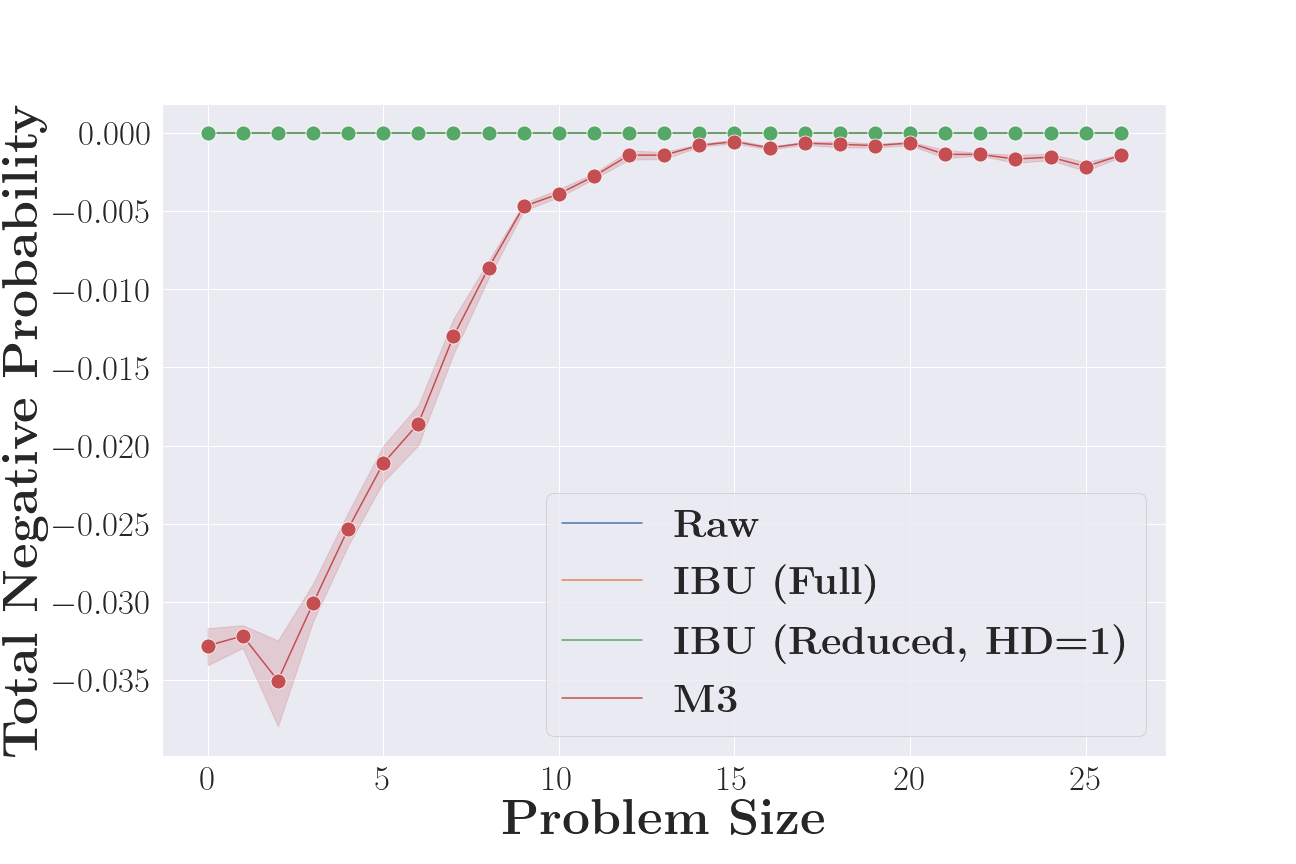}
            \label{fig:weather_filter1}
        \end{subfigure} \\
        \vspace{-11mm}
        \begin{subfigure}{}
            \includegraphics[trim = 0 0 100 70, clip, width=\textwidth]{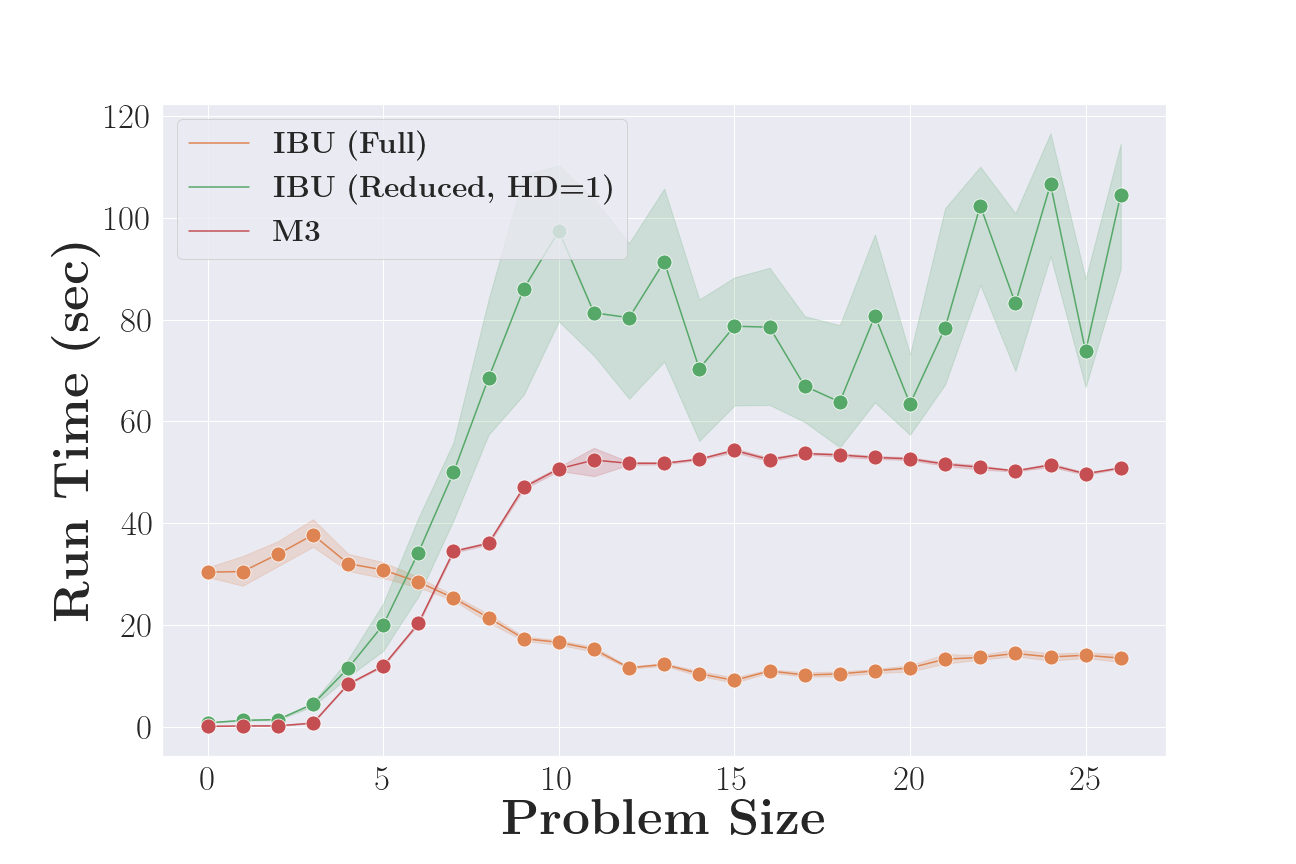}
            \label{fig:weather_filter2}
        \end{subfigure} 
    \end{minipage}
    \caption{\emph{Bernstein-Vazirani algorithm results:} The BV algorithm was implemented on the 27-qubit IBMQ Montreal, and we compare IBU, M3, and the raw (no-MEM) result. The left figure shows log average success probability and is restricted to larger problem sizes as the mitigated and unmitigated distributions are nearly identical at smaller sizes (see Appendix \ref{app:bv}). Full IBU and subspace-reduced IBU have identical performance. The top right figure shows the total negative probability in the error-mitigated distribution, and the bottom right figure shows our machine's run-time. Error bars indicate 95\% confidence intervals.} \label{fig:bv}
\end{figure*}

\paragraph{Problem Setting}  BV~\cite{bernsteinQuantumComplexityTheory1997} is an oracular algorithm where a hidden bitstring $b \in \{ 0,1 \}^n $ is known to the oracle, but when enquired with a known bitstring $x \in \{ 0,1 \}^n $ the oracle reveals $x \oplus b$, where $\oplus$ is bit-wise addition. The goal is to guess the hidden bitstring $b$ by making queries to the oracle. Classically, this algorithm requires $O(n)$ queries, but the BV algorithm solves this problem with $O(1)$ query. Theoretically, regardless of the problem size or the hidden bitstring, the success probability for BV is always 1 (see \cref{fig:bv_circuit}). This expectation is violated for a real quantum computer which is subject to decoherence. The data, originally collected by Ref.~\cite{pokharelDemonstrationAlgorithmicQuantum2022}, was taken from a publicly available repository. This work demonstrated quantum speedup over the classical strategy for single-shot BV, where the hidden bitstring changes after every oracle query. While the original work computed time-to-solution as the success metric, here we focus on how the raw success probabilities are affected by MEM. 

\paragraph{Experimental Details} In Ref.~\cite{pokharelDemonstrationAlgorithmicQuantum2022}, the BV experiment was performed on the 27-qubit IBMQ Montreal. Experiments at each problem size had 32000 shots, and we use $\nicefrac{1}{32000}$ as the convergence tolerance for IBU. The primary metric is the success probability, i.e., the probability that a single guess yields the right answer. This problem employs up to 26 qubits on a bonafide quantum algorithm, and we explored whether MEM can improve this success probability. 

\paragraph{Results} Our results are shown in \cref{fig:bv} (additional results in \cref{app:bv}), and we see that a 
significant improvement in the success probability is possible due to MEM. 
In terms of the success probabilities, our IBU implementation performs about equal to or marginally better than the state-of-the-art M3 method~\cite{nationScalableMitigationMeasurement2021}. Notably, the distribution produced by IBU does not contain negative probabilities, while negative probabilities are present in the quasi-probability solutions provided by M3 (see \cref{fig:bv}). In terms of runtime, as the problem size increases, full IBU implementation, i.e., the one without subspace reduction, is noticeably faster than M3. In contrast, the subspace-reduced version, which is set to track bitstrings up to Hamming distance 1, is slightly slower. Recall that this is because compared to full IBU, subspace-reduced IBU is less memory intensive but can be somewhat slower. For a smaller number of qubits, relying on full IBU is reasonable, but the memory constraints become the bottleneck for larger problem sizes.

\begin{figure*}[htpb!]
    \centering
    \begin{minipage}{0.6\linewidth}
            \begin{subfigure}{}
                \includegraphics[trim = 50 0 100 100, clip,width=\textwidth]{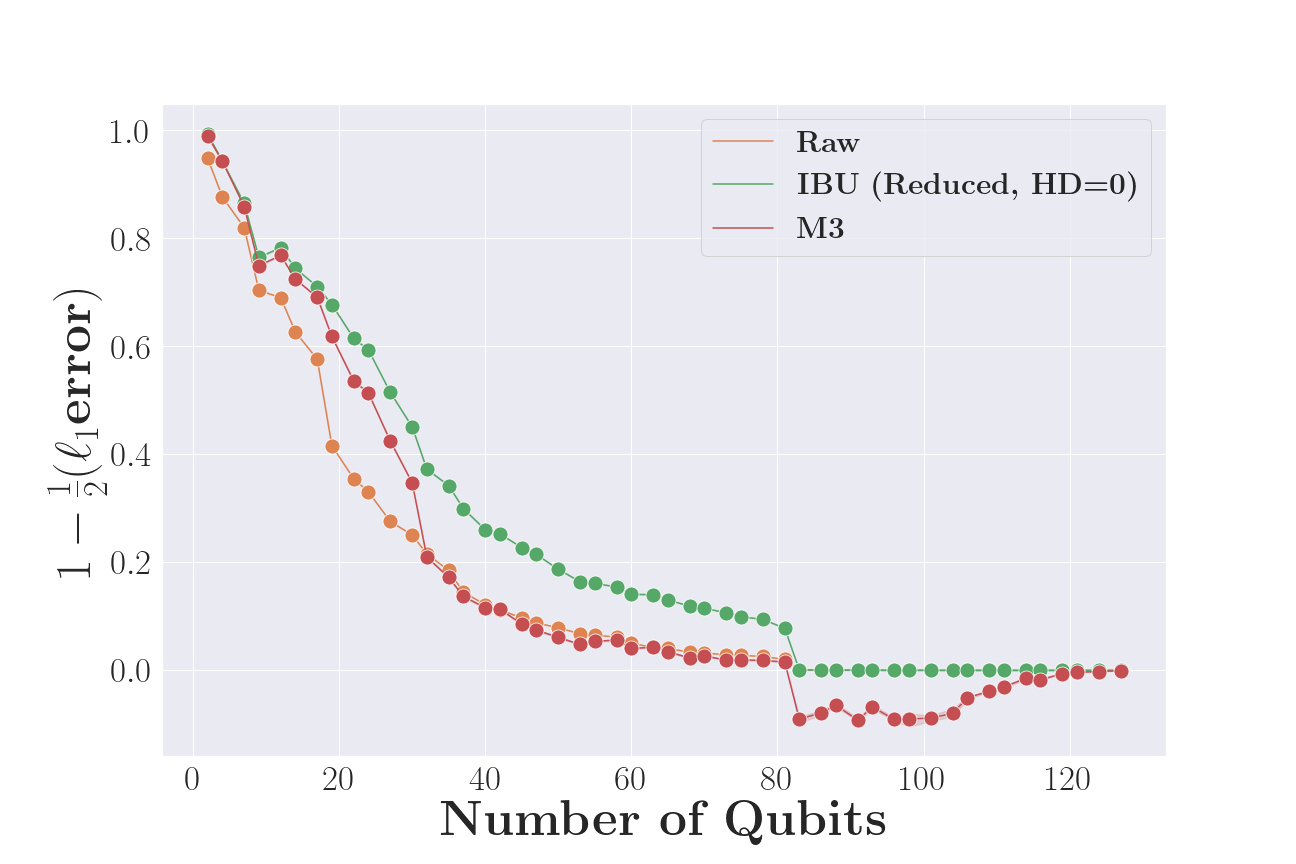}
            \end{subfigure}
        \end{minipage}
    \begin{minipage}{0.38\linewidth}
        \begin{subfigure}{}
            \includegraphics[trim = 0 70 100 0, clip,width=\textwidth]{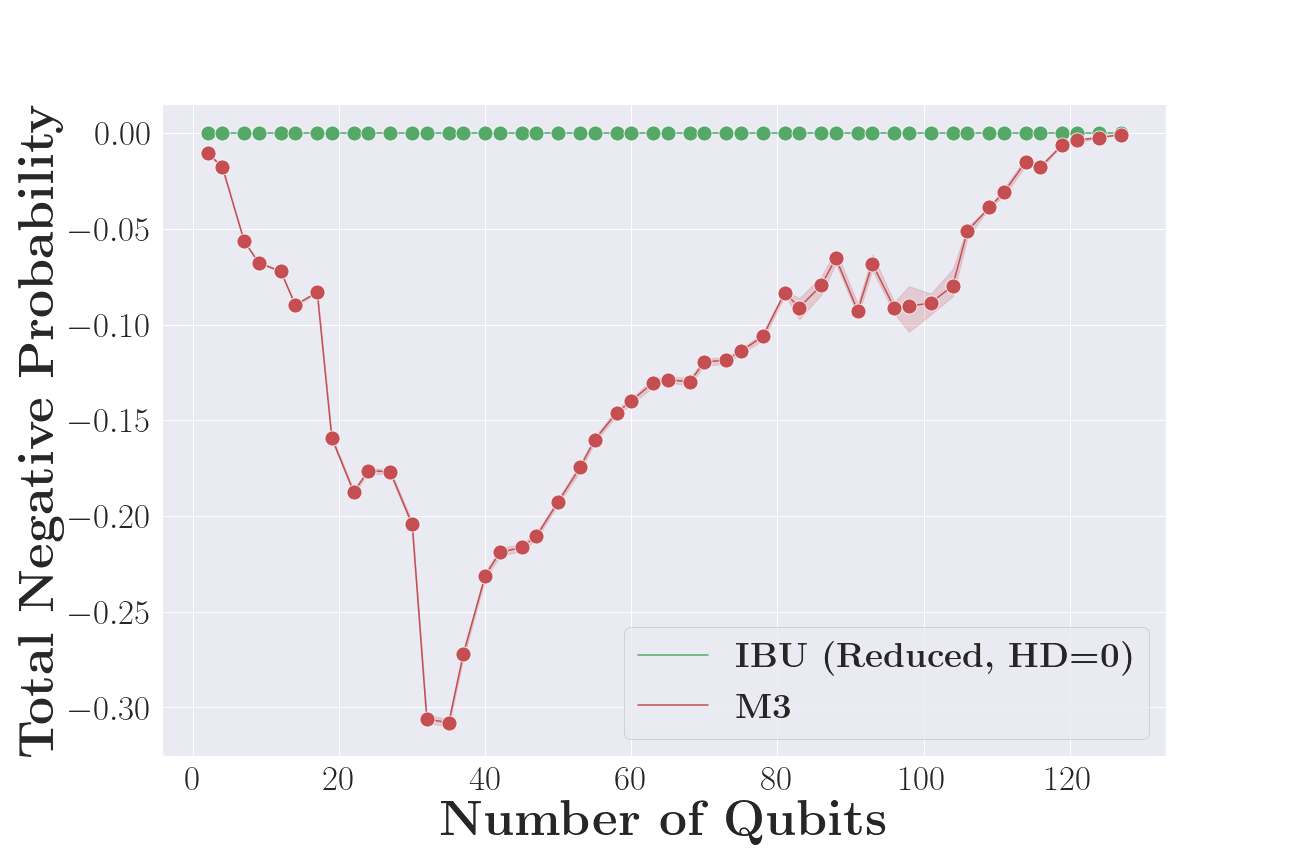}
        \end{subfigure} \\
                \vspace{-7.3mm}
        \begin{subfigure}{}
            \includegraphics[trim = 0 0 100 70, clip,width=\textwidth]{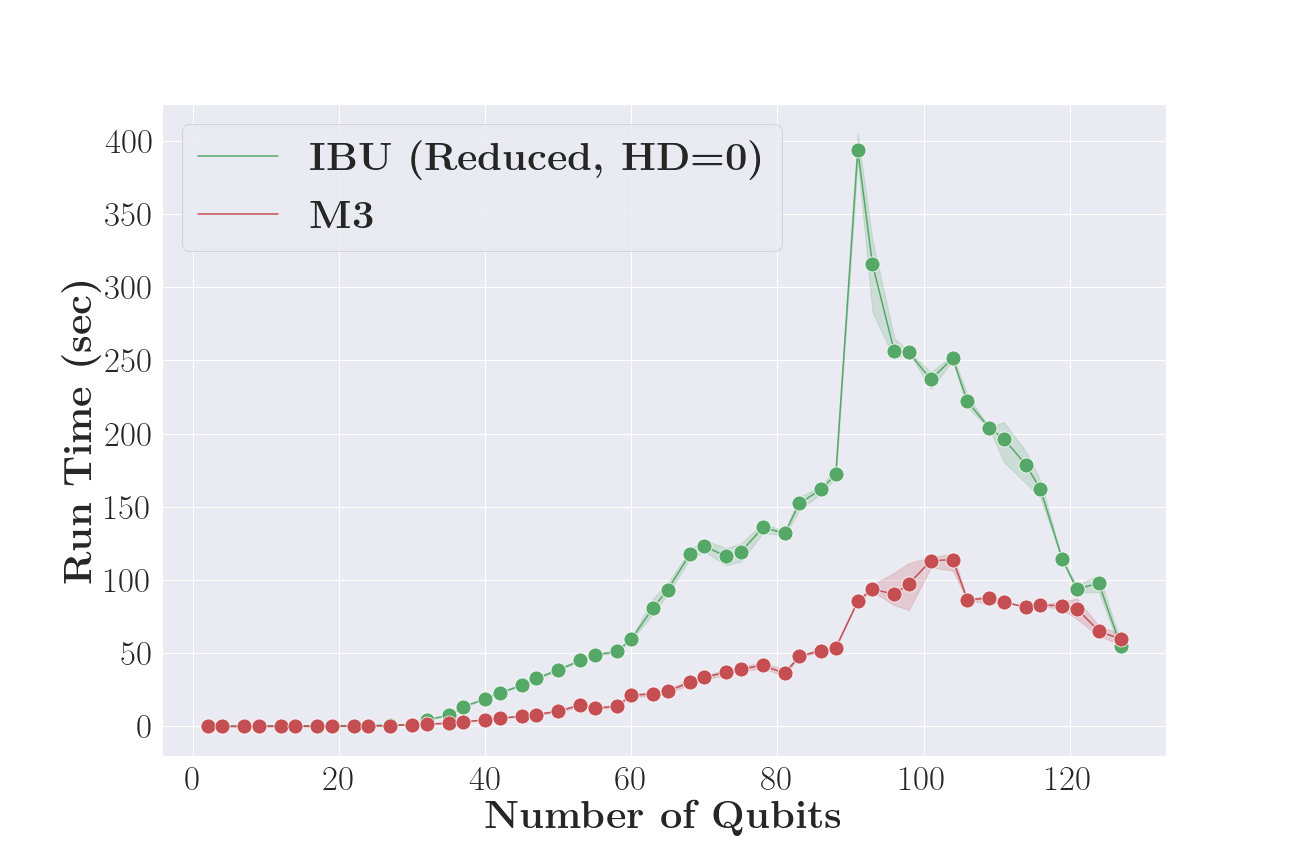}
        \end{subfigure} 
    \end{minipage}
    \caption{\emph{GHZ state preparation on 127-qubit IBMQ Washington:} The left figure shows the normalized $\ell_1$ score. We visualize results for the full range of GHZ states we attempted to create, up to 127 qubits. Circuit errors dominate beyond $n=81$, and MEM is not of much help. At all problem sizes, IBU returns a higher value for the performance metric $1- \frac12 (\ell_1 \text{ error})$. M3 achieves a negative score due to negative probabilities. The top right figure shows the total negative probability in the error-mitigated distribution; here, M3 yields significant negative probabilities.  The bottom right figure shows the run time on our machine. IBU takes longer to run than M3 but is reasonable in absolute terms. 
    }
    \label{fig:ghz-washington}
\end{figure*}

\subsection{Generating GHZ states}
\label{sec:ghz}

\paragraph{Data Generation} In this experiment, we prepared the GHZ state, $| \text{GHZ}(n) \rangle = \frac{| 0 \rangle^{\otimes n} + | 1 \rangle^{\otimes n} }{ \sqrt{2} }$ on the 127-qubit ibmq\_washington (see Appendix \ref{sec:dev} for device architecture). GHZ states are prepared by entangling $n$ qubits. This means that as long as there is a single connection allowing for a multi-qubit entangling operation like CNOT, it is possible to create a GHZ state linearly increasing in circuit depth. To reduce circuit-depth, the entangling operation is implemented by parallelizing the two-qubit operations as much as the device architecture allows. Starting from the most connected qubit (the qubit with the greatest number of connections/edges to neighboring qubits) in a superposition state $|+\rangle$, we use breadth-first search to entangle every additional qubit with CNOT gates. In effect, this amounts to prioritizing the distance $1$ neighbors to entangle, followed by the distance $2$ neighbors, and so on. We give an example of this strategy in \cref{fig:ghz-prep}.

\paragraph{Experimental Details} We generated GHZ states up to $n=127$.  The experiments here repeated for 100,000 shots, and the results were bootstrapped using the observed counts. We use a convergence tolerance of $\nicefrac{1}{10000}$. The primary error metric is a \emph{normalized $\ell_1$ score}; the $\ell_1$ error between two probability distributions $\vec{p}, \vec{q}$ is $\sum_i |{p}_i - {q}_i|_1$ and lies in $[0, 2]$, and we normalize it as $1 - \frac12(\ell_1$ error) to lie in the range $[0, 1]$ where a score of 1 implies zero $\ell_1$ error. We consider the $\ell_1$ error $=\sum_i |{\theta}_i - {q}_i|_1$ between the error mitigated distribution $\vec{\theta}$ and the theoretical distribution $\vec{q}$ of bitstrings for a GHZ state measured in the $Z$ basis (i.e., $\vec{q} = [0.5, 0, 0, \ldots, 0, 0.5]$). Due to the memory costs, we can only run IBU with subspace reduction. Furthermore, we use a Hamming distance of zero, meaning MEM modifies the probability mass only on the measured bitstrings.





\paragraph{Results}    \cref{fig:ghz-washington} shows our GHZ state preparation experiment results. Firstly, IBU has a significantly higher normalized $\ell_1$ score than the raw data or M3. 
Beyond $n=81$, the true solution bitstrings ($0^n$ and $1^n$) were no longer the modal bitstrings due to circuit errors, so MEM (M3 and IBU) is not of much help. However, the normalized $\ell_1$ score for IBU goes to zero, but the score for M3 is negative due to negative probabilities. We found that the negative probabilities produced by M3 can be quite substantial, up to $\sim 30\%$ of the probability mass. We also found that M3 places higher probability mass on the `correct' bitstrings ($0^n$ and $1^n$), but compensates for this with negative probabilities on the wrong bitstrings, leading to overall worse performance on the normalized $\ell_1$ score (see Appendix \ref{sec:addresibm}). Finally, we note that the cost of achieving better $\ell_1$ error is longer run-time: IBU is noticeably slower than M3. Nevertheless, in absolute terms, IBU can perform MEM on 81 qubits in a little over 2 minutes on our machine. IBU run-time declines past $n=81$ as iterations do not modify the guess, and there is little room for progress. One can always choose a higher tolerance to reduce the number of IBU iterations and hence run-time. In Appendix \ref{sec:addresibm}, we also provide a graph showing empirically that the per-iteration run-time scales linearly in the number of qubits.


%








\section{Conclusion}
\label{sec:conclusion}

We showed that IBU, an expectation-maximization algorithm, can be implemented scalably and without compromising on the benefits of error mitigation. Our implementation is scalable to QC with hundreds of qubits while avoiding negative probabilities. The mitigation accuracy and runtime are as good and often better than the state-of-the-art MEM methods. We demonstrated our method by mitigating results from the 26-qubit BV algorithm and 127-qubit GHZ preparation experiment. Using an $\ell_1$-based metric, we showed that up to 81 qubits, the GHZ experiment had a non-zero overlap with the ideal GHZ probability distribution. To our knowledge, this is the largest number of qubits to be entangled on programmable QCs. Overall, our results highlight the need to mitigate readout errors and provide an efficient way to do so. 

As we stress above, MEM strategies rely on certain physical assumptions about measurement errors. In our case, we assume linear errors with minimal measurement crosstalk. The runtime and efficacy of various strategies depend on the validity of these assumptions. As QCs become more sophisticated, frequently validating these assumptions will be necessary. In the future, we expect such sophisticated quantum detector tomography to be native to the QC operation and inform MEM strategies.

\section{Acknowledgements}
This work was partly supported by the U.S. Department of Energy, Office of Science, Office of Advanced Scientific Computing Research, and Accelerated Research in Quantum Computing under Award Number DE-SC0020316. This research used resources from the Oak Ridge Leadership Computing Facility, which is a DOE Office of Science User Facility supported under Contract DE-AC05-00OR22725.

\newpage 

\phantom{=}
\newpage 

\bibliography{biblio.bib, mem.bib}

\newpage 

\phantom{=}
\newpage

\appendix


\section{Iterative Bayesian Unfolding as Expectation-Maximization}\label{sec:fullderiv}

We give the complete derivation showing that iterative Bayesian unfolding for MEM is an instantiation of the well-known \emph{Expectation-Maximization} (EM) algorithm in machine learning with multinomial distributions. 



\paragraph{Why EM}
To see why EM is needed, observe that the likelihood of observing our data given a choice of parameter $\vec{\theta}$ can be written as:
\begin{equation}
        P( {\bf Z'}; \vec{\theta}) = \prod_{i=1}^S \sum_{j=1}^{2^n} P(z'_i | z_j) P_\theta(z_j)
\end{equation}
Although we can take the log-likelihood to eliminate the product, we are left an irreducible sum inside the log that makes it difficult to analytically solve for the optimal paramters:
\begin{equation}\label{eq:loglik}
        \ell(\vec{\theta}; {\bf Z'}) =  \log P( {\bf Z'}; \vec{\theta}) = \sum_{i=1}^S \log \left(\sum_{j=1}^{2^n} P(z'_i | z_j) P_\theta(z_j) \right)
\end{equation}

EM is designed to get around this issue by instead maximizing a lower bound of the log-likelihood $\ell(\vec{\theta}; {\bf Z'})$.

\paragraph{Expectation-Maximization} We begin by observing that if we knew the parameters $\vec{\theta}$, we could easily compute the joint probability of any observed bitstring and the `true' bitstring (i.e., the bitstring that would have been measured except for measurement error) as $P(z'_i, z_j; \vec{\theta}) = P(z'_i|z_j)P_\theta(z_j)$. At the same time, if we knew the joint probability $P(z'_i, z_j; \vec{\theta})$ of any observed bitstring and any given latent bitstring, we would be able to invert the previous equation to infer parameters $\vec{\theta}$. EM turns this logic into an iterative algorithm that alternates between computing the conditional expected joint log-likelihood of observed and latent bitstrings given parameters $\vec{\theta}^k$ and finding the parameter $\vec{\theta}^{k+1}$ that maximizes this conditional expected log-likelihood. 
By Jensen's inequality, we can manipulate Eq \ref{eq:loglik} as:
\begin{equation}\small \label{eq:bound}
\begin{split}
    \ell(\vec{\theta}; {\bf Z'}) &= \sum_{i=1}^S \log \left(\sum_{j=1}^{2^n} P(z'_i | z_j) P_\theta(z_j) \right) \\
    &= \sum_{i=1}^S \log \left(\sum_{j=1}^{2^n} P(z'_i | z_j) P_\theta(z_j) \frac{P_{\theta^k}(z_j|z'_i)}{P_{\theta^k}(z_j|z'_i)} \right) \\
    &= \sum_{i=1}^S \log \left(\sum_{j=1}^{2^n} P_{\theta^k}(z_j|z'_i) \frac{P(z'_i | z_j) P_\theta(z_j)}{P_{\theta^k}(z_j|z'_i)} \right) \\
    &\geq \sum_{i=1}^S \sum_{j=1}^{2^n} P_{\theta^k}(z_j|z'_i) \log \left( \frac{P(z'_i, z_j; \vec{\theta})}{P_{\theta^k}(z_j|z'_i)} \right) \\
    &\equiv \hat{\ell}(\vec{\theta} | \vec{\theta}^k; {\bf Z'})
    \end{split}
\end{equation}

We can maximize this lower bound of the log-likelihood $\hat{\ell}(\vec{\theta} | \vec{\theta}^k; {\bf Z'})$ locally by updating the parameter at a given time-step $\vec{\theta}^k$ as follows:

\begin{equation}
    \begin{split}
        \vec{\theta}^{k+1} &= \underset{\vec{\theta}}{\text{arg max }}\sum_{i=1}^S \sum_{j=1}^{2^n} P_{\theta^k}(z_j|z'_i) \log \left( \frac{P(z'_i, z_j; \vec{\theta})}{P_{\theta^k}(z_j|z'_i)} \right)  \\
        &= \underset{\vec{\theta}}{\text{arg max }} \sum_{i=1}^S \sum_{j=1}^{2^n} P_{\theta^k}(z_j|z'_i) \log P(z'_i, z_j; \vec{\theta}) 
    \end{split}
\end{equation}

Define $\ell(\vec{\theta} | \vec{\theta}^k; {\bf Z'}) := \sum_{i=1}^S \sum_{j=1}^{2^n} P_{\theta^k}(z_j|z'_i) \log P(z'_i, z_j; \vec{\theta}) $, and note that maximizing $\hat{\ell}(\vec{\theta} | \vec{\theta}^k; {\bf Z'})$ is equivalent to maximizing $\ell(\vec{\theta} | \vec{\theta}^k; {\bf Z'})$.  EM consists of an \emph{E}-step which involves computing $\ell(\vec{\theta} | \vec{\theta}^k; {\bf Z'})$ and an \emph{M}-step which involves computing $\text{arg max}_{\vec{\theta}} \ell ( \vec{\theta} | \vec{\theta}^k ; \mathbf{Z'})$. For computational reasons, we will combine these into a single step for updating parameter $\vec{\theta}^k$ to $\vec{\theta}^{k+1}$. To locally maximize $\ell(\vec{\theta} | \vec{\theta}^k; {\bf Z'})$, we can take the derivative (using the Lagrange multiplier $\lambda$ to enforce the normalization constraint that $\sum_j \vec{\theta}_j = 1$).

\begin{equation} \small
    \begin{split}
        &\phantom{=} \frac{\partial }{ \partial {\theta}_j} \left( \sum_{i=1}^S \sum_{j'=1}^{2^n} P_{\theta^k}(z_{j'}|z'_i) \log P(z'_i, z_{j'}; \vec{\theta})  + \lambda \left(1 - \sum_{j'=1}^{2^n} {\theta}_{j'} \right) \right)\\
        &= \frac{\partial }{ \partial {\theta}_j} \left( \sum_{i=1}^S \sum_{j'=1}^{2^n} P_{\theta^k}(z_{j'}|z'_i) \log P_\theta(z_{j'}) + \lambda \left(1 - \sum_{j'=1}^{2^n} {\theta}_{j'} \right) \right) \\
        &= \frac{\partial }{ \partial {\theta}_j} \left( \sum_{i=1}^S \sum_{j'=1}^{2^n} P_{\theta^k}(z_{j'}|z'_i) \log {\theta}_{j'} + \lambda \left(1 - \sum_{j'=1}^{2^n} {\theta}_{j'} \right)  \right) \\
        &= \left(\sum_{i=1}^S P_{\theta^k}(z_j|z'_i) \right)\frac1{{\theta}_j} - \lambda
    \end{split}
\end{equation}

Setting this to zero, we get ${\theta}_j = \frac1{\lambda}\left(\sum_{i=1}^S P_{\theta^k}(z_j|z'_i) \right)$, and by the requirement that $\sum_j {\theta}_j = 1 \Rightarrow \sum_j \frac1{\lambda}\left(\sum_{i=1}^S P_{\theta^k}(z_j|z'_i) \right) = 1 \Rightarrow \lambda =  \sum_{i=1}^S \sum_j P_{\theta^k}(z_j|z'_i) = \sum_{i=1}^S 1 = S$. Substituting this back in, we get our update rule for each element ${\theta}_j$ of $\vec{\theta}$ that locally maximizes the log-likelihood given guess $\vec{\theta}^k$:

\begin{equation}
    {\theta}^{k+1}_j = \frac1S \sum_{i=1}^S P_{\theta^k}(z_j|z'_i)
\end{equation}

Note that by Bayes' rule, we have $ P_{\theta^k}(z_j|z'_i) = \frac{P(z'_i | z_j) P_{\theta^k}(z_j)}{\sum_m P(z'_i | z_m) P_{\theta^k}(z_m)}$. Additionally, instead of summing over every bitstring in the dataset, we can re-write the sum as over all possible bitstrings using bitstring counts $c(z'_i)$. Doing so, we end up with the IBU update rule:

\begin{equation}\label{eq:ibu2}
    {\theta}^{k+1}_j = \sum_{i=1}^{2^n} \frac{c(z'_i)}S  \cdot \frac{P(z'_i | z_j) P_{\theta^k}(z_j)}{\sum_m P(z'_i | z_m) P_{\theta^k}(z_m)}
\end{equation}


\paragraph{Convergence of EM}  From Equation \ref{eq:bound}, we know that $\hat{\ell}(\vec{\theta} | \vec{\theta}^k; {\bf Z'})$ lower bounds the log-likelihood of the data ${\ell}(\vec{\theta}; {\bf Z'})$, i.e., that ${\ell}(\vec{\theta}; {\bf Z'}) \geq \hat{\ell}(\vec{\theta} | \vec{\theta}^k; {\bf Z'})$. We can further show that these two functions coincide at the current choice of parameter values $\vec{\theta} = \vec{\theta}^k$:
\begin{equation}
\begin{split}
    \hat{\ell}(\vec{\theta}^k | \vec{\theta}^k; {\bf Z'}) &= \sum_{i=1}^S \sum_{j=1}^{2^n} P_{\theta^k}(z_j|z'_i) \log \left( \frac{P(z'_i, z_j; \vec{\theta}^k)}{P_{\theta^k}(z_j|z'_i)} \right) \\
    &= \sum_{i=1}^S \sum_{j=1}^{2^n} P_{\theta^k}(z_j|z'_i) \log P(z'_i; \vec{\theta}^k) \\
    &= \sum_{i=1}^S \log P(z'_i; \vec{\theta}^k) \sum_{j=1}^{2^n} P_{\theta^k}(z_j|z'_i)  \\
    &= \sum_{i=1}^S \log P(z'_i; \vec{\theta}^k) \cdot 1 \\
    &= \ell(\vec{\theta}^k; {\bf Z'})
\end{split}
\end{equation}

Together, these facts imply that any choice of $\vec{\theta}^{k+1}$ which improves $\hat{\ell}(\vec{\theta}^{k+1} | \vec{\theta}^k; {\bf Z'})$ from the current estimate $\hat{\ell}(\vec{\theta}^k | \vec{\theta}^k; {\bf Z'})$ must also improve the log-likelihood, i.e., $\hat{\ell}(\vec{\theta}^{k+1} | \vec{\theta}^k; {\bf Z'}) > \hat{\ell}(\vec{\theta}^k | \vec{\theta}^k; {\bf Z'}) \Rightarrow \ell(\vec{\theta}^{k+1}; {\bf Z'}) > \ell(\vec{\theta}^k; {\bf Z'})$ since $\ell(\vec{\theta}^{k+1}; {\bf Z'})  \geq \hat{\ell}(\vec{\theta}^{k+1} | \vec{\theta}^k; {\bf Z'}) > \hat{\ell}(\vec{\theta}^k | \vec{\theta}^k; {\bf Z'}) = \ell(\vec{\theta}^k; {\bf Z'}) $. Since EM updates $\vec{\theta}^k$ to maximize $ \hat{\ell}(\vec{\theta} | \vec{\theta}^k; {\bf Z'})$ locally, the log-likelihood $\ell(\vec{\theta}^k; {\bf Z'})$ increases at every iteration. Indeed, the log-likelihood increases by at least the amount $\hat{\ell}(\vec{\theta}^{k+1} | \vec{\theta}^k; {\bf Z'}) - \hat{\ell}(\vec{\theta}^k | \vec{\theta}^k; {\bf Z'})$. The algorithm iterates until it reaches a stationary point of $\hat{\ell}(\vec{\theta} | \vec{\theta}^k; {\bf Z'})$. Since $\hat{\ell}(\vec{\theta} | \vec{\theta}^k; {\bf Z'})$ is tangent to the log-likelihood $\ell(\vec{\theta}; {\bf Z'})$ at $\vec{\theta} = \vec{\theta}^k$ \citep{bishop2006pattern}, the termination also coincides with a stationary point of the log-likelihood. Thus, EM converges to a local stationary point of the log-likelihood.

\subsection{MAP IBU}

Here we suggest a simple Bayesian extension of IBU: allowing us to impose a prior on the parameters $P(\vec{\theta})$ and identifying the Maximum a posteriori (MAP) estimate of the parameters $P(\vec{\theta} | {\bf Z'})$. We begin by observing that:

\begin{equation}
    \begin{split}
        \log P(\vec{\theta} | {\bf Z'}) &= \log P({\bf Z'} | \vec{\theta}) + \log P(\vec{\theta}) - \log P({\bf Z'})
    \end{split}
\end{equation}

Since we wish to find the choice of $\vec{\theta}$ at which the posterior is maximized, we ignore $\log P({\bf Z'})$. Further, by Equation \ref{eq:bound}, we know that $\log P({\bf Z'} | \vec{\theta}) \geq \hat{\ell}(\vec{\theta}|\vec{\theta}^k; {\bf Z'})$. Similar to the previous approach, instead of directly maximizing the posterior, we seek to maximize the lower-bound:

\begin{equation}
    \vec{\theta}^{k+1} = \underset{\vec{\theta}}{\text{arg max } } \hat{\ell}(\vec{\theta}|\vec{\theta}^k; {\bf Z'}) + \log P(\vec{\theta})
\end{equation}

For our purposes, we assume that the prior over parameters is specified as a Dirichlet distribution with parameter $\boldsymbol{\alpha} \in \mathds{R}^{2^n}$, and let $K= \sum_j \alpha_j$. This is a natural choice as Dirichlet distributions are the conjugate prior for multinomial distribution. Thus, $P(\vec{\theta}) = \frac1{B(\boldsymbol{\alpha})} \prod_{j=1}^{2^n} {\theta}_j^{\alpha_j - 1}$ where $B(\cdot)$ is the beta function. Taking the derivative:

\begin{equation}
    \begin{split}
        &\phantom{=} \frac{\partial }{ \partial {\theta}_j} \left( \hat{\ell}(\vec{\theta}|\vec{\theta}^k; {\bf Z'}) + \log P(\vec{\theta}) + \lambda \left(1 - \sum_{j'=1}^{2^n} {\theta}_{j'} \right)\right) \\
        &= \left(\sum_{i=1}^S P_{\theta^k}(z_j|z'_i) \right)\frac1{{\theta}_j} - \lambda + \frac{\partial }{ \partial {\theta}_j} \left( \log P(\vec{\theta})\right) \\
        &= \left(\sum_{i=1}^S P_{\theta^k}(z_j|z'_i) \right)\frac1{{\theta}_j} - \lambda + \frac{\partial }{ \partial {\theta}_j} \log \left(  \frac1{B(\boldsymbol{\alpha})} \prod_{j'=1}^{2^n} {\theta}_{j'}^{\alpha_{j'} - 1}\right) \\
        &= \left(\sum_{i=1}^S P_{\theta^k}(z_j|z'_i) \right)\frac1{{\theta}_j} - \lambda + \frac{\partial }{ \partial {\theta}_j} \sum_{j'=1}^{2^n} ({\alpha_{j'} - 1}) \log \left(   {\theta}_{j'}\right) \\
        &= \left(\sum_{i=1}^S P_{\theta^k}(z_j|z'_i) \right)\frac1{{\theta}_j} - \lambda + \frac{\alpha_j - 1}{{\theta}_j}
    \end{split}
\end{equation}

Setting this to zero, we get ${\theta}_j = \frac1{\lambda}\left((\alpha_j - 1) + \sum_{i=1}^S P_{\theta^k}(z_j|z'_i) \right)$, and by the requirement that: 

\begin{equation}
\begin{split}
    &\phantom{=} \sum_{j=1}^{2^n} {\theta}_j = 1 \\
    &\Rightarrow \sum_{j=1}^{2^n} \frac1{\lambda}\left((\alpha_j - 1) +\sum_{i=1}^S P_{\theta^k}(z_j|z'_i) \right) = 1 \\
    &\Rightarrow \lambda = \sum_{j=1}^{2^n} (\alpha_j - 1) +  \sum_{i=1}^S \sum_j P_{\theta^k}(z_j|z'_i) \\
    &\Rightarrow \lambda = S + K - 2^n
\end{split}
\end{equation}

Substituting this back in, we get our update rule for each element ${\theta}_j$ of $\vec{\theta}$ that locally maximizes the log-likelihood given an initial guess $\vec{\theta}^k$:

\begin{equation}
    {\theta}^{k+1}_j =  \frac1{S + K - 2^n}\left((\alpha_j - 1) + \sum_{i=1}^S P_{\theta^k}(z_j|z'_i) \right)
\end{equation}

As before, we can rewrite the sum with  counts $c(z'_i)$ and use Bayes' rule to arrive at the update rule:

\begin{equation}\label{eq:mapibu}
    {\theta}^{k+1}_j = \frac{\alpha_j - 1}{S+K-2^n} + \sum_{i=1}^{2^n} \frac{c(z'_i)}{S + K - 2^n}  \cdot \frac{P(z'_i | z_j) P_{\theta^k}(z_j)}{\sum_m P(z'_i | z_m) P_{\theta^k}(z_m)}
\end{equation}

If we impose a uniform prior by setting $\alpha_j = 1$ for all $j$, then Equation \ref{eq:mapibu} clearly reduces to standard IBU as in Equation \ref{eq:ibu2}. When using non-uniform priors, MAP-IBU prevents the parameters from updating too far away from the specified prior.

\begin{figure*}[htpb!]
    \includegraphics[ width=0.6\textwidth]{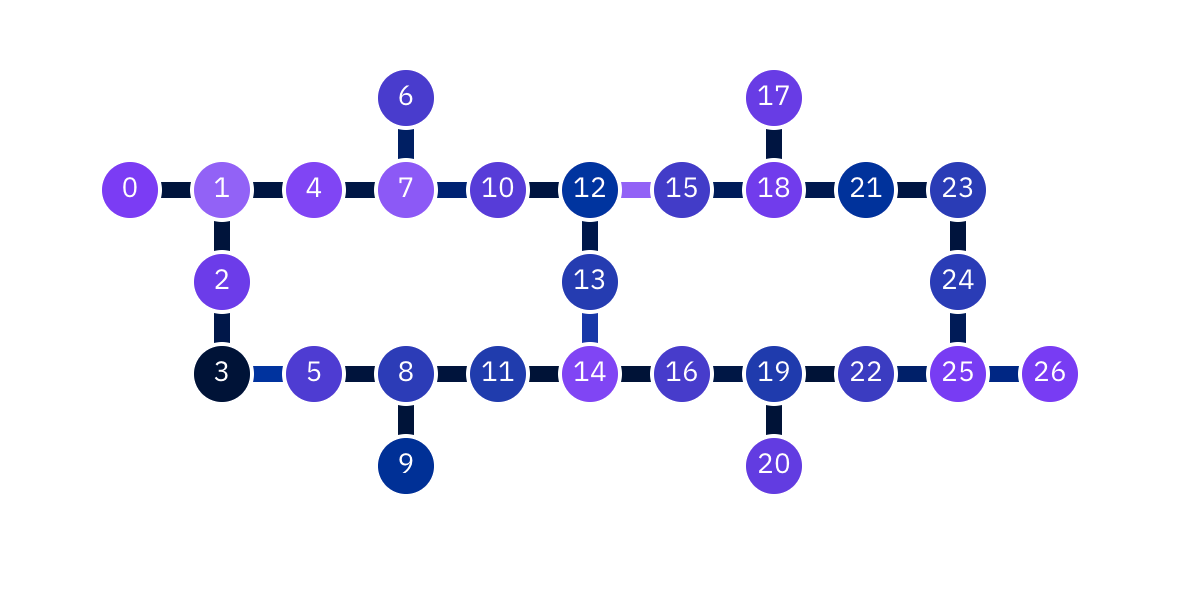}
    \caption{\emph{Device connectivity for IBMQ Montreal.} The entire device was used to implement 26-qubit Bernstein-Vazirani algorithm. 
    \label{fig:montreal-connectivity}}
\end{figure*} 

\begin{figure*}[htpb!]
    \includegraphics[width=0.7\textwidth]{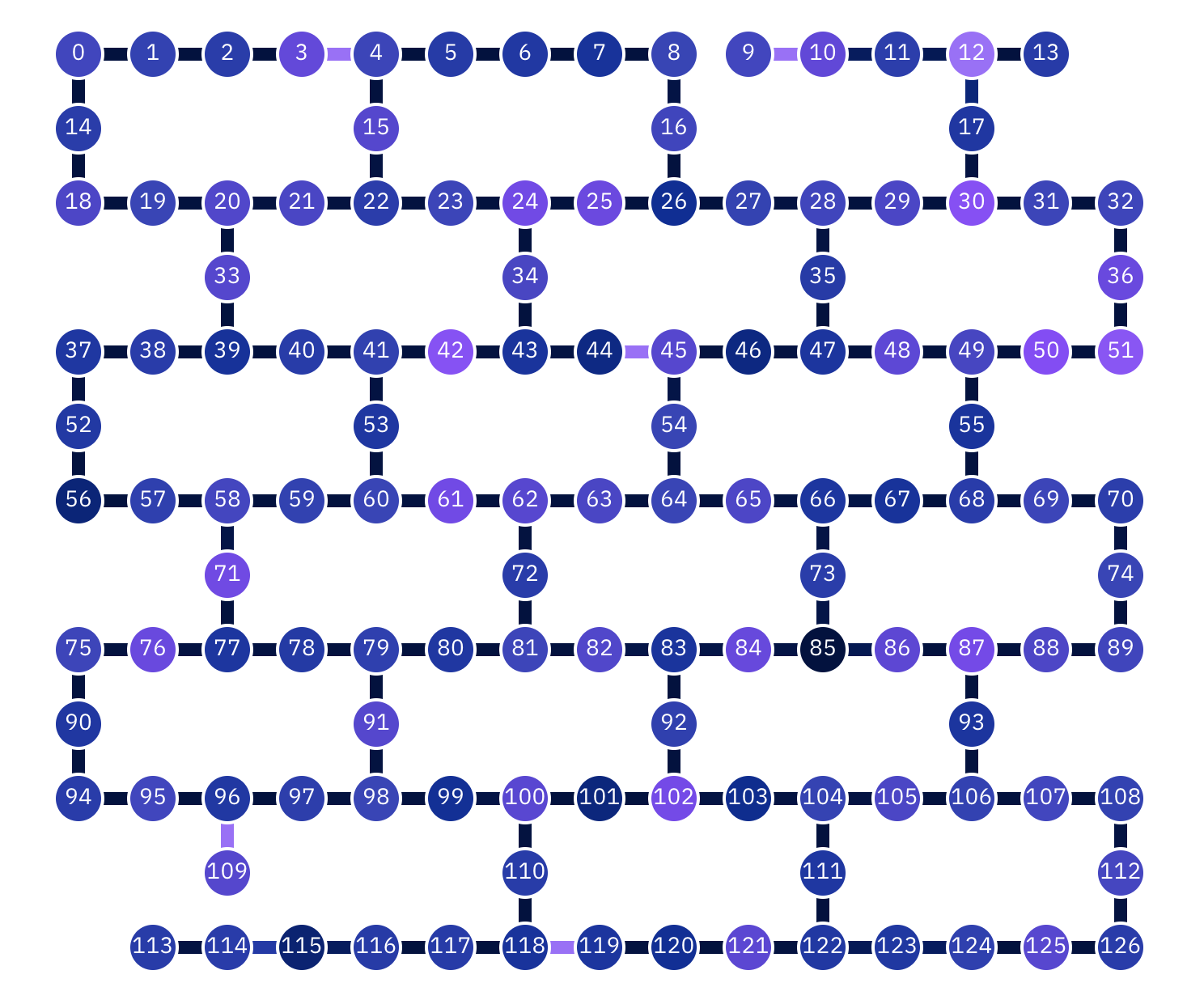}
    \caption{\emph{Device connectivity for IBMQ Washington.} We constructed GHZ$(n)$ states for $n=2$ to $n=127$. However, the `correct' bitstrings (all 0s and all 1s) were no longer the modal measured bitstrings beyond $n=81$ qubits, and the normalized $\ell_1$ score was 0 (see \cref{fig:ghz-washington}). 
    \label{fig:washington-connectivity}}    
\end{figure*}

\section{Device Specifications}\label{sec:dev}

IBMQ Montreal and IBMQ Washington are 27-qubit and 127-qubit devices built using the IBMQE Falcon r5.11 and Eagle r1 processors, respectively~\cite{IBMQuantum2022}. They have quantum volumes of 128 and 64. We show the device architectures in Figures \ref{fig:montreal-connectivity} and \ref{fig:washington-connectivity}. \cref{tab:device} shows the gate errors, readout errors and the $T_1$ and $T_2$ times for both devices.  

\begin{table}[h!]
    \centering
    \begin{tabular}{|l|c|c|c|c|c|c|}
    \hline
        ~ & \multicolumn{2}{c}{\bf Montreal}  & & \multicolumn{2}{c}{\bf Washington} &  \\ \hline
        ~ & Min & Mean & Max & Min & Mean & Max \\ \hline
        $T_1$ ($\mu$s) & 57.57 & 113.2 & 187.1 & 40.43 & 101.98 & 247.7 \\ \hline
        $T_2$ ($\mu$s) & 22.58 & 99.72 & 198.71 & 2.61 & 94.29 & 259.39\\ \hline
        1QG Error  (\%) & 0.02 & 0.04 & 0.15 & 0.02 & 0.08 & 1.09 \\ \hline
        2QG Error  (\%) & 0.57 & 1.35 & 6.09 &  0.53 & 4.79 & 100.   \\ \hline
        1QG Duration ($\mu$s) & 0.04 & 0.04 & 0.04 & 0.04 & 0.04 & 0.04 \\ \hline
        2QG Duration ($\mu$s) & 0.27 & 0.43 & 0.63 & 0.27 & 0.55 & 1.34 \\ \hline
        RO Error (\%) & 0.79 & 2.59 & 10.66 & 0.23  & 3.51 & 38.49 \\ \hline
        RO Duration  ($\mu$s) & 5.2 & 5.2 & 5.2 & 0.86 & 0.86 & 0.86 \\ \hline
    \end{tabular}
\caption{\emph{Device specifications for Montreal and Washington.} 1QG and 2QG denote 1-qubit gate and 2-qubit gate, respectively. RO denotes readout.}
\label{tab:device}
\end{table}

\subsection{Additional Details for Bernstein-Vazirani experiment}\label{app:bv}

\begin{figure*}[hptb!]
    \centering
    \begin{minipage}{0.49\linewidth}
            \begin{subfigure}{}
                \includegraphics[trim = 50 0 100 100, clip,width=\textwidth]{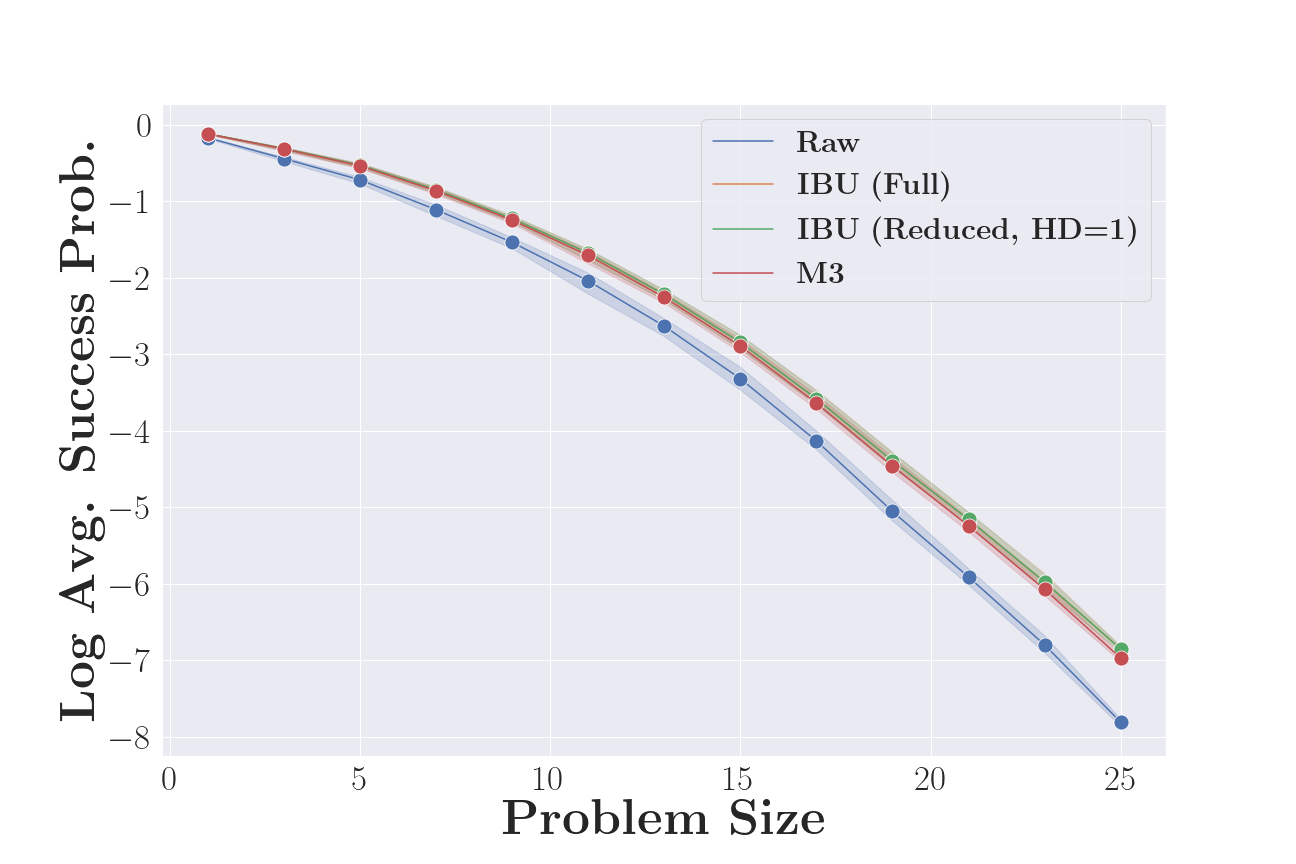}
            \end{subfigure}
        \end{minipage}
    \begin{minipage}{0.49\linewidth}
        \begin{subfigure}{}
            \includegraphics[trim = 50 0 100 100, clip,width=\textwidth]{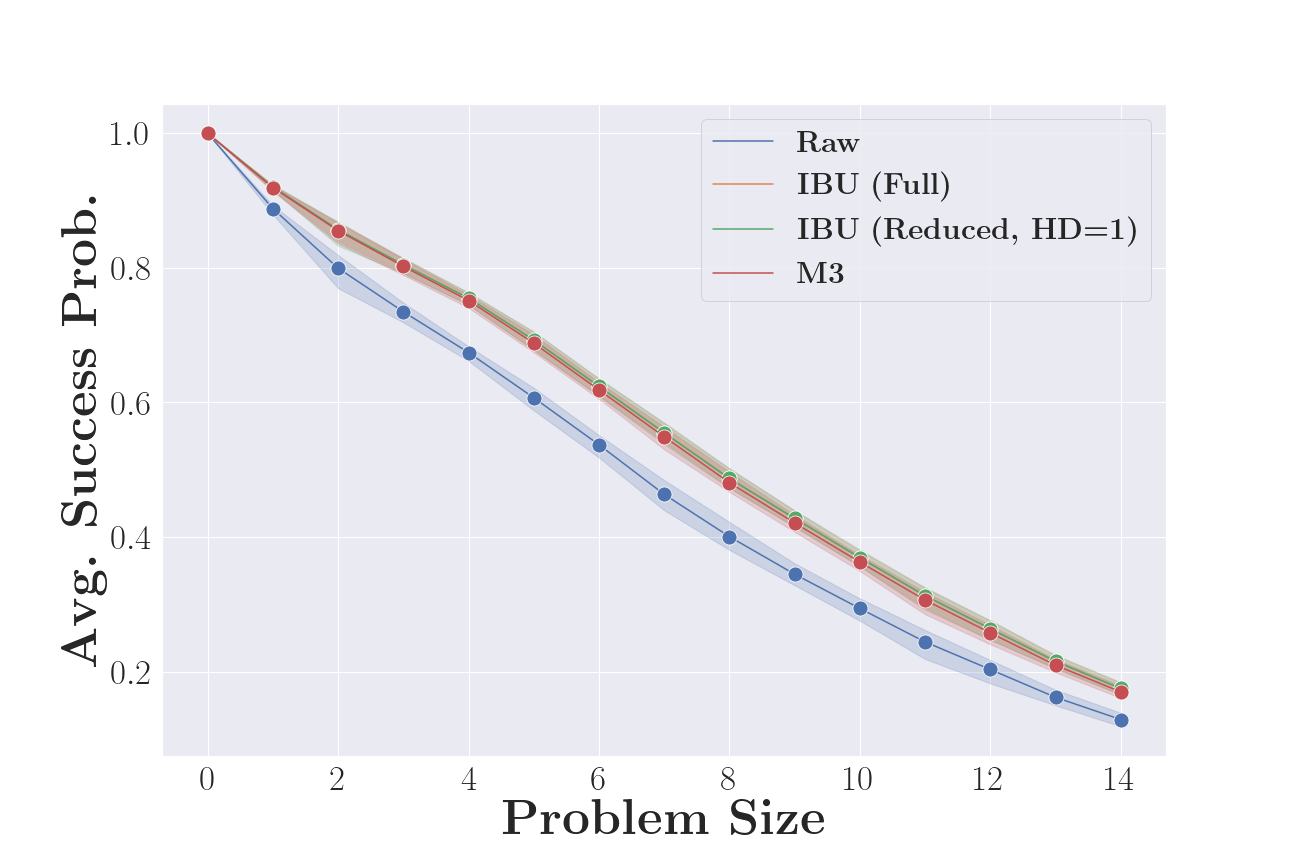}
        \end{subfigure}
    \end{minipage}
    \caption{\emph{Additional Bernstein-Vazirani results:} Bernstein-Vazirani results for all problem sizes. The left figure shows the log average success probability for the full range of problem sizes from 1 to 26. The figure on the right shows the average success probability for small problem sizes. The performance of IBU (both full and reduced subspace) and M3 are nearly identical at all problem sizes. Error bars indicate 95\% confidence intervals.
    }
    \label{fig:bv-extra}
\end{figure*}

In \cref{fig:bv-extra}, we show the log average success probability for all problem sizes from 1 to 26 and the average success probability (not on the log scale) for problem sizes ranging from 1 to 14. The latter provides small-sized problems. Our scalable IBU implementation is effective at small sizes as well. 


\subsection{Additional Results for GHZ state preparation experiment}\label{sec:addresibm}

\begin{figure*}[hptb!]
     \includegraphics[width=\textwidth]{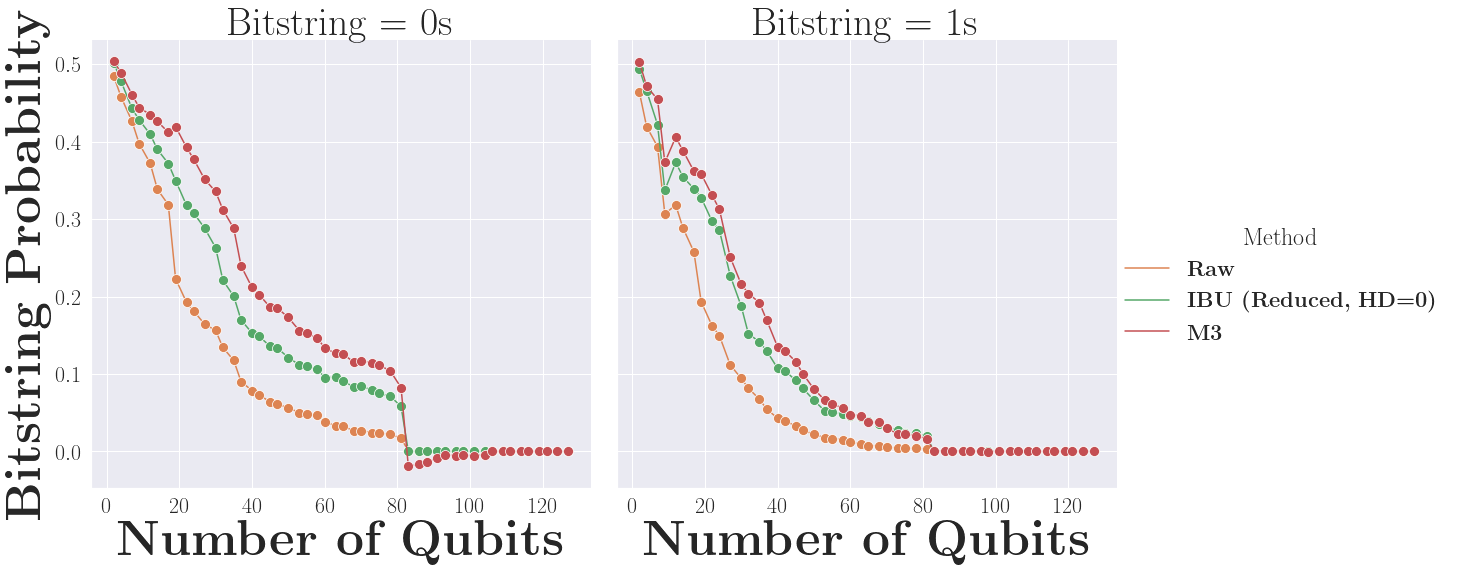}
    \caption{\emph{GHZ state preparation on 127-qubit IBMQ Washington}: The probabilities assigned to the `correct' bitstrings (`$0^n$' on the left and `$1^n$' on the right) with IBU, M3, and no MEM. Even though IBU achieved a higher normalized $\ell_1$ score (\cref{fig:ghz-washington}), M3 places greater probability on the `correct' bitstrings. This is because M3 compensates for higher probabilities on the correct answer by assigning negative probabilities to `wrong' bitstrings.} 
    \label{fig:ghz-washington3}
\end{figure*}

In \cref{fig:ghz-washington3}, we visualize the probabilities assigned to the `correct' bitstrings (`$0^n$' and `$1^n$') with IBU, M3, and no MEM. Even though IBU achieved a higher normalized $\ell_1$ score (\cref{fig:ghz-washington}), M3 places greater probability on the `correct' bitstrings and compensates for this with negative probabilities on `wrong' bitstrings. While this may appear advantageous, the negative probabilities show that M3's error-mitigated distribution deviates from the true solution, which assigns zero probability to these bitstrings. Hence, the normalized $\ell_1$ score gives a more reliable measure of whether the error-mitigated distribution matches the true distribution. Indeed, \cref{fig:ghz-washington3} we see that M3 assigns negative probabilities to the `correct' bitstring (`$0^n$') at $n>81$.

\begin{figure*}[hptb!]
    \includegraphics[trim = 0 0 100 70, clip, width=\textwidth]{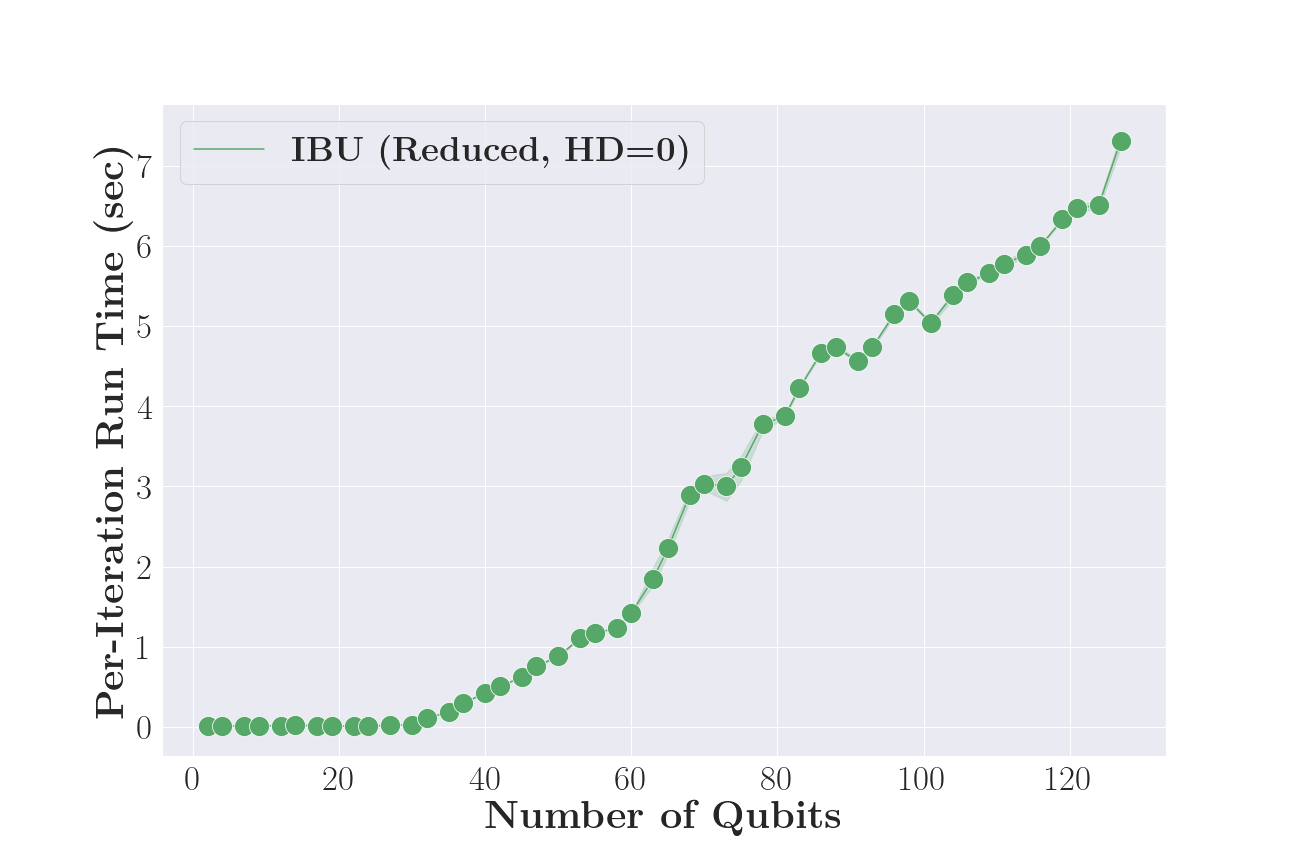}
    \caption{\emph{GHZ state preparation on 127-qubit IBMQ Washington (Per-Iteration Run-time)}: To disregard the effect of a varying number of iterations till convergence, We give the time taken by a single IBU iteration under Hamming distance 0. The run-time scales linearly with the number of qubits, when using Hamming distance 0, and the number of shots, matching theoretical expectations and showing that IBU is scalable to a higher number of qubits. Error bars indicate 95\% confidence intervals.}
    \label{fig:ibmrt2}
\end{figure*}

In \cref{fig:ibmrt2}, we show the per-iteration runtime of IBU. This disregards the effect that the convergence tolerance might have on the run-time. As expected, the graph is essentially linear in the number of qubits.


\end{document}